\documentclass[10pt,twocolumn]{IEEEtran}

\usepackage{amsmath,graphics,amssymb,epsfig,subfigure,color,cite}

\usepackage{amsthm}
\theoremstyle{plain}
\newtheoremstyle{mystyle}
  {0mm}
  {0mm}
  {}
  {4mm}
  {\bfseries}
  {:}
  { }
  {\thmname{#1}\thmnumber{ #2}\thmnote{ (#3)}}
\theoremstyle{mystyle}

\usepackage{array}
\usepackage{multirow}
\usepackage{enumerate}
\usepackage{bm}
\usepackage{float}
\usepackage{makecell}
\usepackage[right=0.625in, left=0.625in, top=0.7in, bottom=1.1in]{geometry}

\usepackage{algorithm}
\usepackage{algpseudocode}
\usepackage{algcompatible}
\algblockdefx{WHILE}{ENDWHILE}[1]%
  {\textbf{while }#1 \textbf{}}%
  {\textbf{end}}
\algblockdefx{FORP}{ENDFORP}[1]%
  {\textbf{for }#1 \textbf{do in parallel}}%
  {\textbf{end}}
\algblockdefx{FOR}{ENDFOR}[1]%
  {\textbf{for }#1 \textbf{do}}%
  {\textbf{end}}
\algblockdefx{IF}{ENDIF}[1]%
  {\textbf{if }#1 \textbf{}}%
  {\textbf{end}} 
\algblockdefx{ELSIF}{ENDIF}[1]%
  {\textbf{else if }#1 }%
  {\textbf{end}}
\algnewcommand\algorithmicprocedure{\textbf{function}}
\algnewcommand\FUNC{\item[\algorithmicprocedure]}%
\algnewcommand\algorithmicendprocedure{\textbf{end function}}
\algnewcommand\ENDFUNC{\item[\algorithmicendprocedure]}%
 \usepackage{setspace}
\let\Algorithm\algorithm
\renewcommand\algorithm[1][]{\Algorithm[#1]\setstretch{1.4}}

\usepackage{bbm}
\usepackage{bigints}
\usepackage{adjustbox}

\floatname{algorithm}{Algorithm}

\usepackage{enumitem}

\newcommand{\argmin}{\operatornamewithlimits{argmin}}
\makeatletter
\newcommand{\vast}{\bBigg@{4.5}}
\newcommand{\Vast}{\bBigg@{7.5}}
\makeatother

\begin{document}
\title{\fontsize{23}{28}\selectfont Blind Training for Channel-Adaptive Digital Semantic Communications
}
\author{Yongjeong Oh, Joohyuk Park, Jinho Choi, Jihong Park, and Yo-Seb Jeon
        \thanks{This work was supported in part by the National Research Foundation of Korea (NRF) funded by Korean Government (MSIT) under Grant RS-2024-00453301 {\em (Corresponding authors: Yo-Seb Jeon and Jihong Park)}.}
	    \thanks{Yongjeong Oh, Joohyuk Park, and Yo-Seb Jeon are with the Department of Electrical Engineering, POSTECH, Pohang, Gyeongbuk 37673, Republic of Korea (email: yongjeongoh@postech.ac.kr, joohyuk.park@postech.ac.kr, yoseb.jeon@postech.ac.kr).}
        \thanks{Jinho Choi is with the School of Electrical and Mechanical Engineering, The University of Adelaide, SA 5005, Australia (email: jinho.choi@adelaide.edu.au).}
        \thanks{Jihong Park is with the Information Systems Technology and Design Pillar, Singapore University of Technology and Design, Singapore 487372 (email: jihong\_park@sutd.edu.sg).}
        }
	\vspace{-2mm}	
	
	\maketitle
	\vspace{-12mm}
\begin{abstract} 

Semantic encoders and decoders for digital semantic communication (SC) often struggle to adapt to variations in unpredictable channel environments and diverse system designs. To address these challenges, this paper proposes a novel framework for training semantic encoders and decoders to enable channel-adaptive digital SC. The core idea is to use binary symmetric channel (BSC)  as a universal representation of generic digital communications, eliminating the need to specify channel environments or system designs. Based on this idea, our framework employs parallel BSCs to equivalently model the relationship between the encoder’s output and the decoder’s input. The bit-flip probabilities of these BSCs are treated as trainable parameters during end-to-end training, with varying levels of regularization applied to address diverse requirements in practical systems. The advantage of our framework is justified by developing a training-aware communication strategy for the inference stage. This strategy makes communication bit errors align with the pre-trained bit-flip probabilities by adaptively selecting power and modulation levels based on practical requirements and channel conditions. Simulation results demonstrate that the proposed framework outperforms existing training approaches in terms of both task performance and power consumption.
\end{abstract}


\section{Introduction}\label{Sec:Intro}

Semantic communication (SC) is a transformative approach to data communication that focuses on conveying the meanings or semantics of raw data relevant to a specific task, rather than accurately transmitting every bit of raw data as in traditional communication \cite{SemCom_Survey_1,SemCom_Survey_2,SemCom_Survey_3,Semantic_Jihong_1,Semantic_Jihong_2}. By sending only essential task-relevant information, it has great potential to enhance bandwidth efficiency and robustness to channel perturbations. A key application is in image and video transmission, where SC extracts semantics through object recognition, scene understanding, and advanced compression algorithms, and transfers perceptually similar content using minimal bandwidth, even under poor channel conditions \cite{SemCom_Image,SemCom_Video_1, SemCom_Video_2}.


Despite its potential, SC remains in its nascent stage, with most of existing frameworks relying on analog-like communication architectures that transmit continuous-valued signals over additive noisy channels \cite{DeepJSCC,DeepJSCC_f,DeepJSCC_Analog_1,DeepJSCC_Analog_2}. This approach is well-suited for deep neural networks (DNNs), which naturally process continuous outputs in their hidden layers, and additive channels can be modeled by a single DNN layer, enabling end-to-end DNN differentiation and trainability. However, this assumption deviates significantly from digital communication systems, limiting the broader adoption and practical deployment of SC. To address this limitation, recent studies \cite{BSC_BEC,NECST,DeepJSCC_Q,JCM,Constellation,JSCC_NOMA,JSCC_Lite,Robust_SNR,JSCC_universal,Joohyuk,AMC_Fixed_Mod} have begun exploring ways to integrate SC into digital communication systems as we will review next.

\subsection{Digital SC: Recent Studies and Limitations}
The transition from analog to digital communication systems presents new challenges for SC, as digital environments require the handling of discrete bits and symbols. To address this, earlier works focused on mapping the real-valued output of the encoder to discrete symbols or bits to improve compatibility with digital communication architectures \cite{BSC_BEC,NECST,DeepJSCC_Q,JCM,Constellation}.
This approach was further extended by incorporating non-orthogonal multiple access to enhance bandwidth efficiency \cite{JSCC_NOMA}, and lightweight DNN structures were developed to accommodate the limited computational power of edge devices \cite{JSCC_Lite}. 

Despite these advancements, digital SC still faces significant challenges, particularly in optimizing power and modulation levels, to achieve high task performance across various communication environments.
This multi-dimensional optimization differs from that of analog SC, which primarily focuses on power optimization alone.  
The primary difficulty in addressing this optimization problem is that the relationship among power, modulation, and task performance is intractable; thus, it is challenging to derive an analytical optimal solution.
Moreover, since the optimal power-modulation pair varies depending on channel conditions, addressing all possible channel conditions, via training either multiple specialized models or a single generalized model, leads to significant training overhead and large amounts of communication data.

Due to the difficulty of determining the optimal power and modulation for varying channel conditions, most digital SC methods train the semantic encoder and decoder in predefined communication environments where the channel distributions, modulation schemes, or transmission power are predetermined
\cite{DeepJSCC_Q,JCM,Constellation,JSCC_NOMA,JSCC_Lite,Robust_SNR,JSCC_universal,Joohyuk}. In addition, it is common to impose unnecessary constraints such as equal transmit power and modulation schemes for all symbols. Since the encoder and decoder trained under these constraints do not generalize well to unexpected environments, their task performance significantly deteriorates with variations in channel conditions, modulation order, and transmit power levels. Furthermore, the strict power and modulation constraints fail to enable prioritization of critical bits or achieve robustness, resulting in task performance degradation.


Recently, to tackle the difficulty in adapting to diverse channel conditions, a model-based training method was proposed \cite{Joohyuk}, in which a one-bit quantizer is applied to the output of the semantic encoder. The quantizer's output is then assumed to undergo statistical channel models, such as the binary symmetric channel (BSC) or the binary symmetric erasure channel (BSEC). 
During the communication stage, the modulation level for each symbol is adjusted based on the channel conditions, while ensuring that the bit error remains below the predefined bit-flip probability used during training. In this strategy, however, the bit-flip probabilities were treated as hyperparameters, which incurs substantial training overhead in hyperparameter tuning. Additionally, due to the difficulty of jointly optimizing power and modulation, only modulation was adjusted, leading to performance limitations. In addition to digital SC methods that consider training, an adaptive modulation method based on pre-trained models was proposed \cite{AMC_Fixed_Mod}. However, this method does not account for the effects of modulation and fading channels during training, so its task performance is inevitably limited. 



\subsection{Contributions}
In this paper, we propose a novel framework for training semantic encoders and decoders in digital SC to address two key challenges: (i) the difficulty of optimizing power and modulation,  and (ii) the challenge of adapting to unpredictable channel conditions and varying system requirements. In the proposed framework, we use a BSC model as a universal representation of generic digital communication processes, allowing us to eliminate the need to specify channel environments or communication system designs. Building on this idea, we jointly train the bit-flip probabilities of parallel BSCs with the semantic encoder-decoder pair in an end-to-end manner. For the inference stage, we design a training-aware communication strategy that ensures communication bit errors align with the pre-trained bit-flip probabilities  by adaptively adjusting power and modulation levels. Simulation results show that our framework outperforms existing training methods that rely on predefined channel environments or system designs.

The major contributions of this paper are summarized as follows:


%

\begin{itemize}
    \item We propose a novel framework for training $K$ representative sets of semantic encoders and decoders that are independent of specific channel environments or communication system designs. Our key idea is to replace the digital communication process between the semantic encoder and decoder with parallel BSCs that have varying bit-flip probabilities. Although a similar approach was suggested in \cite{Joohyuk}, it treated the bit-flip probabilities of the BSCs as hyperparameters, leading to substantial training overhead due to hyperparameter tuning. To overcome this limitation, our framework treats the bit-flip probabilities as trainable parameters and applies a continuous relaxation technique to enable end-to-end training of the bit-flip probabilities alongside the semantic encoder and decoder. Then, we develop a novel one-parameter regularization technique based on a flip vanishing phenomenon. Specifically, in this phenomenon, we found that end-to-end training with BSC modeling inherently favors zero bit-flip probabilities on average, corresponding to an error-free channel. Leveraging this insight, we introduce a regularization term that penalizes training when the average bit-flip probability is small. This enables the encoder-decoder set trained with low penalty to cover good channels and with high penalty to adapt to poor channels. By controlling the degree of penalization using a single parameter, we obtain $K$ representative sets, each covering a distinct average channel condition.

    \item  For the inference stage, we develop a training-aware communication strategy tailored to our blind training framework. In this framework, we select the most suitable encoder-decoder pair based on the given channel conditions under total power and transmission rate constraints, while also determining the appropriate power allocation and modulation scheme for each transmitted symbol. Specifically, we devise two optimization methods to ensure that communication bit errors align with the trained bit-flip probabilities.
    In the first method, we adjust transmission power for each transmission symbol while keeping modulation levels fixed. In the second method, we jointly adjust both transmission power and modulation levels to minimize overall power consumption. Although adaptive modulation has been considered in \cite{JSCC_universal,Joohyuk,AMC_Fixed_Mod}, none of these works explore the joint optimization of both power and modulation levels. This joint optimization provides a significant advantage in reducing overall power consumption while adhering to transmission rate constraints.

    \item  For image transmission tasks with the MNIST \cite{MNIST}, CIFAR-10 \cite{CIFAR10}, and STL-10 \cite{STL10} datasets, our proposed BlindSC achieves up to a 10\% improvement in the peak signal-to-noise ratio (PSNR) and reduces total transmission power by up to 70\%, at an SNR of 20 dB, compared to other digital SC frameworks. Further, the proposed BlindSC achieves the highest structural similarity index measure (SSIM) across varying transmission rate. 
    

\end{itemize}

\begin{figure*}[t]
    \centering
    {\epsfig{file=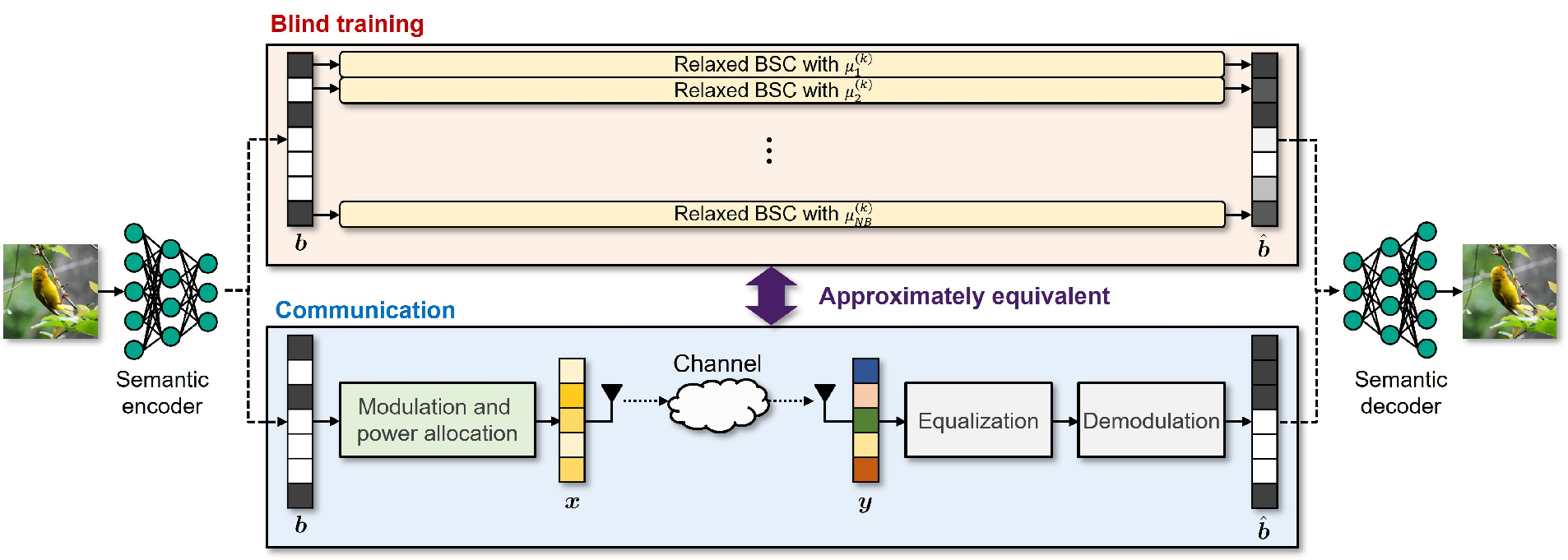,width=14cm}}\vspace{-2mm}
    \caption{The overall structure of BlindSC framework illustrating the approximate equivalence between blind training and actual digital communication systems.}\vspace{-3mm}
    \label{fig:Sys}
\end{figure*}

\section{System Model}\label{Sec:System}

In this work, we consider a typical digital SC system in a wireless network. Specifically, in the system model, we employ a digital deep joint source-channel coding for image transmission \cite{NECST}. 
This system model can be readily extended to support other machine-learning tasks and alternative SC approaches. 

Let ${\bm u} \in \mathbb{R}^M$ be an image data and ${\bm \theta}_{\rm enc}$ be the parameter vector of the encoder. Then, the transmitter encodes the image data as follows:
\begin{align}
    {\bm v} = f_{{\bm \theta}_{\rm enc}}({\bm u}) \in \mathbb{R}^N,
\end{align}
where $f_{{\bm \theta}_{\rm enc}}$ is the encoding function
, and ${\bm v}$ is the semantic feature vector of length $N$. 
After the encoding process, the transmitter quantizes each element of ${\bm v}$ using a $B$-bit quantizer. The resulting quantized value is given by
\begin{align}\label{eq:q_i}
    q_i = {\sf Q}(v_i) \in \mathcal{Q},~i\in\{1,\cdots,N\},
\end{align}
where $v_i$ represents the $i$th element of ${\bm v}$, $\mathcal{Q} = \{\tilde{q}_1,\tilde{q}_2,\cdots,\tilde{q}_{2^B}\}$ is the codebook of the quantizer, and ${\sf Q}: \mathbb{R} \rightarrow \mathcal{Q}$ is the quantization function. For each codeword $\tilde{q}_j$, the corresponding bit sequence is determined as $\tilde{\bm b}_j = {\sf B}(\tilde{q}_i) \in \{0,1\}^{B}$, where ${\sf B}: \mathcal{Q} \rightarrow \{0,1\}^B$ is a mapping function that converts each codeword $\tilde{q}_j$ into a unique binary sequence of length $B$. Based on this, the transmitter converts the quantized value $q_i$ of \eqref{eq:q_i} into its corresponding binary sequence, as follows:
\begin{align}
    {\bm b}_i = {\sf B}(q_i) \in \{0,1\}^{B},~i\in\{1,\cdots,N\}. 
\end{align}
After quantizing all elements of ${\bm v}$, the transmitter obtains the bit sequence ${\bm b} = [{\bm b}_1^{\sf T}, \ldots, {\bm b}_N^{\sf T}]^{\sf T} \in \{0,1\}^{NB}$. Then, through a digital modulation process, this bit sequence is converted into a symbol sequence ${\bm x}$ of length $T$. 

The transmission rate, representing the average number of transmitted bits per channel use, is defined as 
\begin{align}\label{eq:Trans_rate}
    R = \frac{NB}{T},
\end{align}
implying how densely the bits are packed for each channel use. 
To ensure efficient communication, the transmission rate needs to satisfy or exceed a predefined target transmission rate, denoted as $R_{\rm target}$, as follows:
\begin{align}\label{eq:Trans_rate_constraint}
    R \geq R_{\rm target}.
\end{align}
After the modulation process, the transmitter allocates a power $p_{t}$ to each symbol $x_t$ under the total power constraint, which is given by 
\begin{align}\label{eq:Total_power}
    P_{\rm sum}\triangleq \sum_{t=1}^T p_t \leq P_{\rm tot},
\end{align}
where $P_{\rm tot}$ denotes the total available transmission power for transmitting the symbol sequence ${\bm x}$. 

The wireless channel between the transmitter and receiver is modeled as a block fading channel, where the channel coefficients remain constant during the coherence time \cite{goldsmith2005wireless}. Under this assumption, the received signal at time slot $t$ is expressed as
\begin{align} \label{eq:y_t}
    y_t = h\sqrt{p_t}x_t + n_t,~t\in\{1,\ldots,T\},
\end{align}
where $h\in\mathbb{C}$ is a complex-valued channel coefficient, and $n_t \sim \mathcal{CN}(0,\sigma^2)$ represents the additive white Gaussian noise (AWGN). 
The maximum achievable SNR is defined as 
\begin{align}\label{eq:SNR_max}
    {\rm SNR}_{\rm max} = 10\log_{10}\frac{P_{\rm tot}\mathbb{E}[\gamma]}{NB},
\end{align}
where $\gamma=\frac{|h|^2}{\sigma^2}$ is the channel-gain-to-noise-power ratio. 
Assume that the channel coefficient $h$ is accurately estimated at the receiver using pilot-assisted channel estimation during each channel coherence time. Then, the receiver performs channel equalization on the received signal in \eqref{eq:y_t}, resulting in the equalized signal at time slot $t$:
\begin{align} \label{eq:y_t2}
    \tilde{y}_t = \frac{h^*}{|h|^2}y_t = \sqrt{p_t}x_t + \tilde{n}_t,~t\in\{1,\ldots,T\},
\end{align}
where $h^*$ is the complex conjugate of $h$, and $\tilde{n}_t = h^*n_t/|h|^2$. After this, the receiver recovers the estimated symbol sequence $\hat{\bm x}$, followed by demodulation to determine the estimated bit sequence $\hat{\bm b} = [\hat{\bm b}_1^{\sf T}, \ldots, \hat{\bm b}_N^{\sf T}]^{\sf T} \in \{0,1\}^{NB}$.


After reconstructing the bit sequence, the receiver determines the estimated quantized value by performing the inverse process of $\mathcal{B}$, as follows:
\begin{align}
    \hat{q}_i = {\sf B}^{-1}(\hat{\bm b}_i) \in \mathcal{Q},~i\in\{1,\cdots,N\},
\end{align}
where ${\sf B}^{-1}: \{0,1\}^B \rightarrow \mathcal{Q}$. Once the quantized values are obtained, the receiver can reconstruct the original image data using a semantic decoder parameterized by ${\bm \theta}_{\rm dec}$. This decoding operation is denoted as 
\begin{align}
    \hat{\bm u} = f_{{\bm \theta}_{\rm dec}}(\hat{\bm q}) \in \mathbb{R}^K.
\end{align}



In our work, we do not delve into the design and optimization of the quantization-related functions $\mathcal{Q}$, $\mathcal{B}$, and $\mathcal{B}^{-1}$; thereby, we assume that these functions are predefined by employing well-established quantization methods (e.g., \cite{oh2023communication,Quantization,Linear_quant}). In this context, we employ the uniform quantizer defined as $q_i = {\sf Q}(v_i) = \Delta \cdot \left\lfloor \frac{v_i - v_{{\rm min}}}{\Delta} + \frac{1}{2} \right\rfloor + v_{{\rm min}}$, 
where $v_{{\rm min}}$ and $v_{{\rm max}}$ are the minimum and maximum values of quantized values, respectively, and $\Delta=\frac{v_{\rm max}-v_{\rm min}}{2^B-1}$. 
The relationship between $q_i$ and ${\bm b}_i$ is given by $q_i = v_{\rm min} + \Delta\sum_{k=1}^B 2^{k-1}b_i^{(k)}$, where $b_i^{(k)}$ is the $k$th element of ${\bm b}_i$.

Under the system model described above, we propose a novel digital SC framework named BlindSC. Traditional SC methods often struggle to adapt to diverse communication environments because their training requires substantial overhead and large amounts of communication data to create either multiple specialized models or a single generalized model for predefined communication conditions. BlindSC addresses these issues by introducing an error-adaptive blind training strategy, which eliminates the need for prior knowledge of communication factors. Furthermore, it incorporates a training-aware communication strategy that dynamically adjusts transmission parameters based on varying channel conditions. In Sec.~III, we discuss the training strategy of BlindSC in detail, and in Sec.~IV, we explain its communication strategy. 
The high-level procedure of BlindSC is summarized in Fig.~\ref{fig:Sys}.  

\begin{figure*}[t]
    \centering
    {\epsfig{file=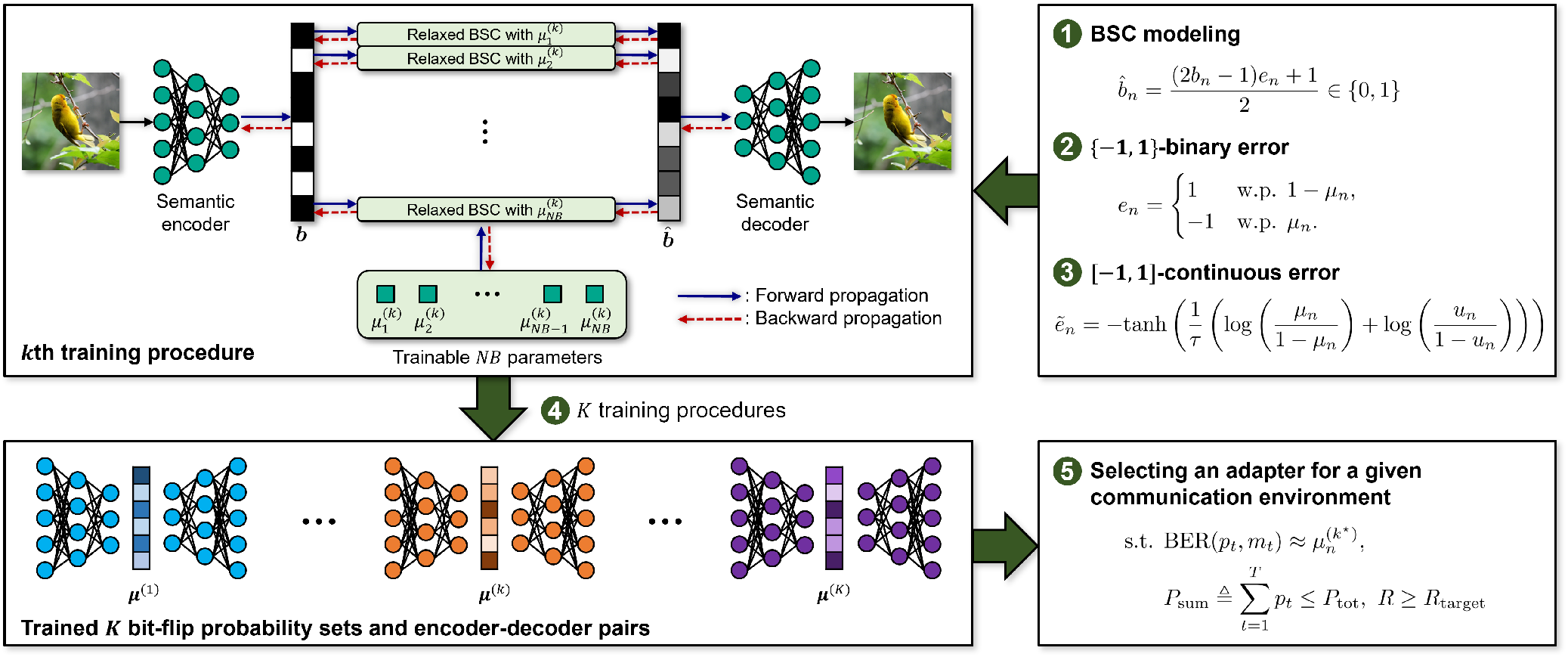,width=15.5cm}}\vspace{-2mm}
    \caption{The proposed blind training strategy in BlindSC with trainable $NB$ parameters and $K$ training procedures.}\vspace{-3mm}
    
    \label{fig:Training}
\end{figure*}

\section{Error-Adaptive Blind Training Strategy\\ of BlindSC}\label{Sec:Comp}
In this section. we introduce an error-adaptive blind training strategy of BlindSC, which allows the semantic encoder and decoder to be trained without prior knowledge of communication factors such as channel conditions, transmission power, and modulation scheme. 




\subsection{BSC Modeling}\label{Sec:BSC}
%


In digital SC, the complex and non-differentiable nature of various communication processes, such as modulation and detection, complicates the training process. To address this, we first characterizes theses processes to a more manageable form by modeling their combined effects using $NB$ BSCs \cite{BSC_BEC,NECST,Joohyuk}. 
It is worth emphasizing that this BSC modeling is used exclusively during training and is not applied in actual communication circumstances, as illustrated in Fig.~\ref{fig:Sys}. Details of the actual communication strategy, including modulation, power allocation, fading channels, and detection, are provided in Sec.~\ref{Sec:infer}.

During training, the relationship between the transmitted $n$th bit $b_n$ and the corresponding received bit $\hat{b}_n$ for the $n$th BSC is modeled as follows:
\begin{align}\label{eq:bsc}
    \hat{b}_n = \frac{(2{b}_n-1)e_n + 1}{2} \in \{0,1\},
\end{align}
where $n\in\{1,\cdots,NB\}$, ${e}_n\in\{-1,1\}$ is a binary error determined as 
\begin{align}\label{eq:error_vec}
    e_n = 
    \begin{cases}
      1, &\text{w.p. }1-\mu_n, \\
      -1, &\text{w.p. }\mu_n,
    \end{cases}
\end{align}
and $\mu_n\in(0,0.5)$ represents the bit-flip probability for the $n$th BSC. 
Note that $\mu_n$ plays a crucial role in characterizing the various processes in digital communication systems. 
For example, if the $n$th bit is transmitted in the $t$th symbol $x_t$ using $2^{m_t}$-QAM with transmission power $p_t$ over a fading channel, 
the bit-flip probability $\mu_n$ for the ML detection can be approximated can be approximated as follows \cite{BER_example}:
\begin{align}\label{eq:mu_example}
    \mu_n \approx& \frac{\sqrt{2^{m_t}} - 1}{\sqrt{2^{m_t}} \log_2 \sqrt{2^{m_t}}} {\rm erfc} \left( \sqrt{\frac{3 {p_t\gamma}}{2(2^{m_t} - 1)}} \right) \nonumber\\ &+ \frac{\sqrt{2^{m_t}} - 2}{\sqrt{2^{m_t}} \log_2 \sqrt{2^{m_t}}} {\rm erfc} \left( 3\sqrt{\frac{3 { p_t\gamma}}{2(2^{m_t} - 1)}} \right),
\end{align}
where ${\rm erfc}(x) = 1 - \frac{2}{\sqrt{\pi}}\int_0^x \exp(-{u^2}){\rm d}u$ is the complementary error function, which is a function of several key factors in digital communication systems, including transmission power $p_t$, channel-gain-to-noise-power ratio $\gamma$, and modulation level $m_t$. 

A key advantage of BSC modeling is its ability to unify the modeling for diverse and unpredictable digital communication systems. This modeling also facilitates end-to-end learning for the semantic encoder and decoder by regarding the linear equation in \eqref{eq:bsc} as a layer of a DNN \cite{BSC_BEC}. 
However, when employing this modeling, it is crucial to determine the bit-flip probabilities $\{\mu_n\}_{n=1}^{NB}$, as they significantly influence the performance of image reconstruction during training. 
One of the primary challenges in determining these bit-flip probabilities is that the relationship between $\mu_n$ and reconstruction performance is intricate and intractable, making it difficult to derive an analytical optimal solution. Most existing digital SC methods have failed to account for this relationship, as their training is typically performed in a specific communication environment, where all symbols are assumed to experience the same bit-flip probability \cite{DeepJSCC_Q,JCM,Constellation,JSCC_NOMA,JSCC_Lite,Robust_SNR,JSCC_universal,Joohyuk}. Additionally, although the concept of BSC-based training was introduced in \cite{BSC_BEC,NECST,Joohyuk}, they still rely on heuristic approaches or fixed communication environments to determine the values of $\{\mu_n\}_{\forall n}$.
To address these challenges, in our training strategy, we introduce the end-to-end learning-based parametric training approach. In this approach, the bit-flip probabilities $\{\mu_n\}_{\forall n}$ are treated as \textit{trainable parameters}, rather than fixed constants predetermined by a specific communication environment. This implies that our training approach eliminates the need for explicit knowledge of communication factors, thereby significantly reducing training overhead and the necessity for extensive communication data collection.  
Further, by incorporating the bit-flip probabilities into the training process, our approach enables their joint optimization alongside the semantic encoder and decoder. 

To further elaborate on our parametric training approach, we start by defining a trainable raw parameter vector, $\tilde{\bm \mu} = [\tilde{\mu}^{\prime}_1,\tilde{\mu}^{\prime}_2,\cdots,{\mu}^{\prime}_{NB}]^{\sf T} \in \mathbb{R}^{NB}$. 
Then, we transform these raw parameters to represent the bit-flip probabilities as follows:
\begin{align}\label{eq:mu}
    \mu_n = \frac{{\rm Sigmoid}({\mu}^{\prime}_n)}{2},~\forall n,
\end{align}
where ${\rm Sigmoid}(x) = \frac{1}{1 + e^{-x}}$ is the sigmoid function. This transformation ensures that the $\mu_n$ falls within the appropriate bit-flip probability range $(0, 0.5)$. 

In the following subsections, we elaborate a more detailed exploration for training the bit-flip probabilities in \eqref{eq:mu}. Specifically, in Sec.~III-B, we focus on deriving the gradient for $\mu_n$ to enable end-to-end learning. Then, in Sec.~III-C, we describe the loss function that guides $\{\mu_n\}_{\forall n}$ to converge to the appropriate values, along with the training method designed to ensure effective adaptation to diverse communication environments. 
The overall procedure of our training strategy is summarized in Fig.~\ref{fig:Training}.

\subsection{End-to-End Learning via Continuous Relaxation}\label{Sec:continuous_relaxation}

The primary challenge in computing the gradient for $\mu_n$ arises from the discrete nature of the BSC described in \eqref{eq:bsc}. Specifically, since the binary error $e_n$ in \eqref{eq:error_vec} follows a \textit{discrete} Bernoulli distribution with $\mu_n$, it is non-trivial to calculate the gradient for $\mu_n$. 
To facilitate the training of $\mu_n$, our training strategy employs the continuous relaxation method in \cite{Continuous_Relax}, which effectively approximates discrete random variables with their continuous counterparts. A representative example of this method is the relaxation of the max function in the Gumbel-max trick to a softmax function, commonly known as the Gumbel-softmax trick. 

Leveraging the continuous relaxation in \cite{Continuous_Relax}, the binary error $e_n$ in \eqref{eq:error_vec} is approximated by its continuous counterpart, given by
\begin{align}\label{eq:approx_error_vec}
    \tilde{e}_n = -{\rm tanh}\left(\frac{1}{\tau}\left(\log \left(\frac{\mu_n}{1-\mu_n}\right) + \log \left(\frac{u_n}{1-u_n}\right)\right)\right),
\end{align}
where $u_1,\ldots,u_{NB}$ are independent and identically distributed random variables, each following the uniform distribution $\mathcal{U}(0,1)$, and $\tau\geq 0$ is the temperature parameter. The detailed derivation of \eqref{eq:approx_error_vec} can be obtained by applying a linear transformation to the Bernoulli random variable described in \cite{Continuous_Relax}. In contrast to the binary error $e_n$, its continuous counterpart $\tilde{e}_n$ is differentiable with respect to $\mu_n$. The corresponding gradient is given by
\begin{align}\label{eq:grad_mu}
    \frac{\partial \tilde{e}_n}{\partial{\mu_n}} =& -{\rm sech}^2\left(  \frac{1}{\tau}\left(\log\left(\frac{\mu_n}{1-\mu_n}\right) + \log\left(\frac{u_n}{1-u_n}\right)\right) \right) \nonumber \\
    & \times \frac{1}{\tau\mu_n(1-\mu_n)},
\end{align}
where ${\rm sech}(x) = \frac{2}{e^x + e^{-x}}$. With this continuous relaxation, substituting $\tilde{e}_n$ into $e_n$ relaxes the input-output relationship of the BSC model in \eqref{eq:bsc} as follows:
\begin{align}\label{eq:bsc_approx}
    \hat{b}_n = \frac{(2{b}_n-1)\tilde{e}_n + 1}{2} \in [0,1]. 
\end{align}


In our training strategy, the relaxed BSC is treated as a layer within the DNN. This integration enables the computation of gradients for the semantic encoder, bit-flip probabilities, and semantic decoder, thereby allowing end-to-end learning for all trainable parameters. Specifically, let $\mathcal{L}$ be the loss function computed at the output of the semantic decoder, which will be detailed in Sec.~III-C. Then, utilizing the chain rule, the gradient of the loss with respect to $\mu_n$ becomes
\begin{align}
    \frac{\partial \mathcal{L}}{\partial \mu_n} = \frac{\partial \mathcal{L}}{\partial \hat{b}_n} \cdot \frac{\partial \hat{b}_n}{\partial \tilde{e}_n} \cdot \frac{\partial \tilde{e}_n}{\partial \mu_n},
\end{align}
where the first component $\frac{\partial \mathcal{L}}{\partial \hat{b}_n}$ can be obtained by performing backpropagation at the semantic decoder. The second component is obtained from the relaxed BSC, given by 
$\frac{\partial \hat{b}_n}{\partial \tilde{e}_n} = b_n - \frac{1}{2}$. 
Following the similar procedure above, the gradient for the semantic encoder is determined as $\frac{\partial \mathcal{L}}{\partial {\bm \theta}_{\rm enc}} = \sum_{n=1}^{NB} \frac{\partial \mathcal{L}}{\partial \hat{b}_n} \cdot \frac{\partial \hat{b}_n}{\partial b_n} \cdot \frac{\partial b_n}{\partial {\bm \theta}_{\rm enc}}$
where the second term is derived from the relaxed BSC, given by $\frac{\partial \hat{b}_n}{\partial b_n} = \tilde{e}_n$. For other trainable parameters, the gradients $\frac{\partial \mathcal{L}}{\partial \mu_n^\prime}$ and $\frac{\partial \mathcal{L}}{\partial {\bm \theta}_{\rm dec}}$, are readily computed by performing backpropagation with the chain rule. 
Meanwhile, in practice, these gradients can be computed using automatic differentiation techniques provided by modern deep learning frameworks, such as TensorFlow and Pytorch. Therefore, our gradient computation is not overly complex, enabling ease of implementation and efficient optimization. 

\subsection{Regularization for Optimizing $\{\mu_n\}_{\forall n}$ in End-to-End Learning}\label{Sec:loss}


The design of the loss function $\mathcal{L}$ is crucial for jointly training $\{\mu_n\}_{\forall n}$ along with the semantic encoder and decoder. However, due to the large number of $\{\mu_n\}_{\forall n}$, it is challenging to control each one individually and to construct an appropriate loss function. To address this, we propose a novel one-parameter regularization technique based on a flip vanishing phenomenon. This phenomenon indicates that end-to-end training with BSC modeling favors zero bit-flip probabilities on average, particularly when the loss function is designed to minimize the image reconstruction error only, such as MSE. This is because lower average bit-flip probabilities naturally lead to fewer transmission errors, thereby reducing the reconstruction error.

Leveraging this insight, we introduce a regularization term that penalizes training when the mean bit-flip probability is small, as follows:
\begin{align}\label{eq:loss}
    \mathcal{L} = \mathbb{E}_{{\bm u},\hat{\bm u}}[d({\bm u},\hat{\bm u})] + \lambda \mathcal{R}({\bm \mu}),
\end{align}
where $d({\bm u},\hat{\bm u})$ is a distortion measure between the original input image ${\bm u}$ and its reconstruction $\hat{\bm u}$, $\lambda > 0$ is a regularization weight that controls the strength of the regularization, and $\mathcal{R}({\bm \mu})$ represents a regularization term that encourages the parameter $\mu_n$ to converge to larger values by penalizing it when its value is small. 
A representative example of our loss function is a combination of MSE and L2-based regularization, formulated as
\begin{align}\label{eq:mse_l2}
    \mathcal{L} = \frac{1}{M}\mathbb{E}_{{\bm u},\hat{\bm u}}\left[\|{\bm u} - \hat{\bm u}\|^2\right] +  \frac{\lambda}{NB}\sum_{n=1}^{NB}\left(\frac{1}{2}-{\mu}_n\right)^2,
\end{align}
where the second term, representing L2 regularization, penalizes deviations of $\{\mu_n\}_{\forall n}$ from their maximum value $0.5$.

In our loss function of \eqref{eq:loss}, increasing $\lambda$ causes the average value of ${\bm \mu}$ to converge toward higher values due to the greater significance of the regularization term. This increase, however, is attained at the cost of increased distortion, as the training procedure places more emphasis on accommodating challenging communication environments rather than minimizing distortion. Therefore, our training strategy involves a trade-off between accommodating challenging communication environments and achieving high task performance.

Leveraging the trade-off above, our key idea for effectively adapting to diverse communication environments is to represent these environments using $K$ distinct bit-flip probability sets.
Let ${\bm \mu}^{(k)}$ be the $k$th bit-flip probability set, where $k\in\{1,\cdots,K\}$. Then, based on the loss function in \eqref{eq:loss}, we perform $K$ training procedures, each with a different regularization strength, $\lambda_k,~k\in\{1,\cdots,K\}$, such that $\lambda_1 < \cdots < \lambda_K$. Then, we can obtain $K$ different bit-flip probability sets and corresponding encoder-decoder pairs. Here, each bit-flip probability set characterizes a specific communication environment. Specifically, a set with a low average bit-flip probability, trained with a low value of $\lambda_k$, represents favorable communication environments, such as those with high total power or strong channel conditions. On the other hand, a set with a high average bit-flip probability, trained with a high value of $\lambda_k$, corresponds to more challenging communication environments with poor channel conditions or low total power.
Using these trained bit-flip probability sets, the transmitter and receiver can dynamically select the most suitable encoder-decoder pair for the given communication environment. In Sec.~\ref{Sec:infer}, we will provide detailed instructions on how to implement this dynamic selection process. Meanwhile, recall that the bit-flip probability in \eqref{eq:mu_example} encapsulates the several factors in digital communication systems, including channel, power, and modulation levels. As a result, our approach eliminates the need to train a separate model for every possible combination of these parameters, significantly reducing the overall number of required models compared to traditional SC methods.

\vspace{1mm}
{\bf Remark 1 (Design of the Regularization Term $\mathcal{R}(\cdot)$:} 
The design of the regularization term $\mathcal{R}(\cdot)$ plays a crucial role in ensuring that the $K$ bit-flip probability sets $\{{\bm \mu}^{(k)}\}_{k=1}^K$ accommodate various communication environments. One intuitive approach would be to set a target value for the $k$th bit-flip probability set, denoted as $\tilde{\mu}^{(k)}$, and formulate the regularization term as $\mathcal{R}({\bm \mu}^{(k)}) = \sum_{n=1}^{NB}\big(\tilde{\mu}^{(k)} - {\mu}_n^{(k)}\big)^2$, where $\tilde{\mu}^{(k)}$. 
While this target-based regularization provides controllability over ${\mu}_n^{(k)}$'s convergence toward the target value, the relationship between the task performance and the target value is unknown, making the target value a hyperparameter. For this reason, this regularization introduces two hyperparameters, ${\bm \mu}^{(k)}$ and $\lambda_k$. Compared to target-based regularization, the L2-based regularization in \eqref{eq:mse_l2} simplifies the training process by utilizing a single hyperparameter $\lambda_k$. To assess the task performance of both regularizations, we have conducted an experiment on the MNIST dataset under varying $\tilde{\mu}^{(1)}$ and $\lambda_1$ values with $K=1$. Table~\ref{table:L2_target} shows that after hyperparameter tuning, both the L2-based and target-based regularizations achieve comparable PSNR performance. However, the L2-based regularization achieves this performance using only one hyperparameter, while the target-based regularization requires tuning two parameters. This result implies that optimizing a single hyperparameter, as in the L2-based regularization, appears to be sufficient to achieve high task performance across various communication environments.

\begin{table}[t]
    \renewcommand{\arraystretch}{1.05}
    \caption{Comparsion of the PSNRs for task-based and L2-based regularizations across varying $\tilde{\mu}^{(1)}$ and $\lambda_1$ values at ${\rm SNR}_{\rm max}\in\{10,20\}$~${\rm d}$B on the MNIST dataset.}\label{table:L2_target}
    \centering
    \setlength{\tabcolsep}{4pt}
    \footnotesize
    \begin{tabular}{|wc{1.8cm}|wc{0.5cm}|wc{0.8cm}|wc{1.2cm}|wc{1.2cm}|wc{1.2cm}|}
        \hline
        \multicolumn{3}{|c|}{\multirow{2}{*}{${\rm SNR}_{\rm max}=10$ dB}} & \multicolumn{3}{c|}{$\lambda_1$} \\ \cline{4-6}
        \multicolumn{3}{|c|}{} & $10^{-4}$ & $10^{-3}$ & $10^{-2}$ \\ \hline
        \multirow{3}{*}{\begin{tabular}[c]{@{}c@{}}Target-based\\ regularization\end{tabular}} & \multirow{3}{*}{$\tilde{\mu}^{(1)}$} & $0.1$ & $34.47$ & $35.06$ & ${\bf 36.08}$ \\ \cline{3-6}
        & & $0.2$ & $34.41$ & $35.74$ & ${\bf 36.08}$ \\ \cline{3-6}
        & & $0.3$ & $35.05$ & $35.99$ & $35.49$ \\ \hline
        \multicolumn{3}{|c|}{L2-based regularization} & $35.31$ & ${\bf 36.12}$ & $33.74$  \\ \hline \hline
        \multicolumn{3}{|c|}{\multirow{2}{*}{${\rm SNR}_{\rm max}=20$ dB}} & \multicolumn{3}{c|}{$\lambda_1$} \\ \cline{4-6}
        \multicolumn{3}{|c|}{} & $10^{-4}$ & $10^{-3}$ & $10^{-2}$ \\ \hline
        \multirow{3}{*}{\begin{tabular}[c]{@{}c@{}}Target-based\\ regularization\end{tabular}} & \multirow{3}{*}{$\tilde{\mu}^{(1)}$} & $0.1$ & $42.03$ & $42.04$ & $41.32$ \\ \cline{3-6}
        & & $0.2$ & $41.93$ & $41.82$ & $39.69$ \\ \cline{3-6}
        & & $0.3$ & ${\bf 42.11}$ & $41.45$ & $38.25$ \\ \hline
        \multicolumn{3}{|c|}{L2-based regularization} & ${\bf 42.09}$ & $40.70$ & $35.56$ \\ \hline
    \end{tabular}
\end{table}

\begin{figure}[t]
    \centering
    {\epsfig{file=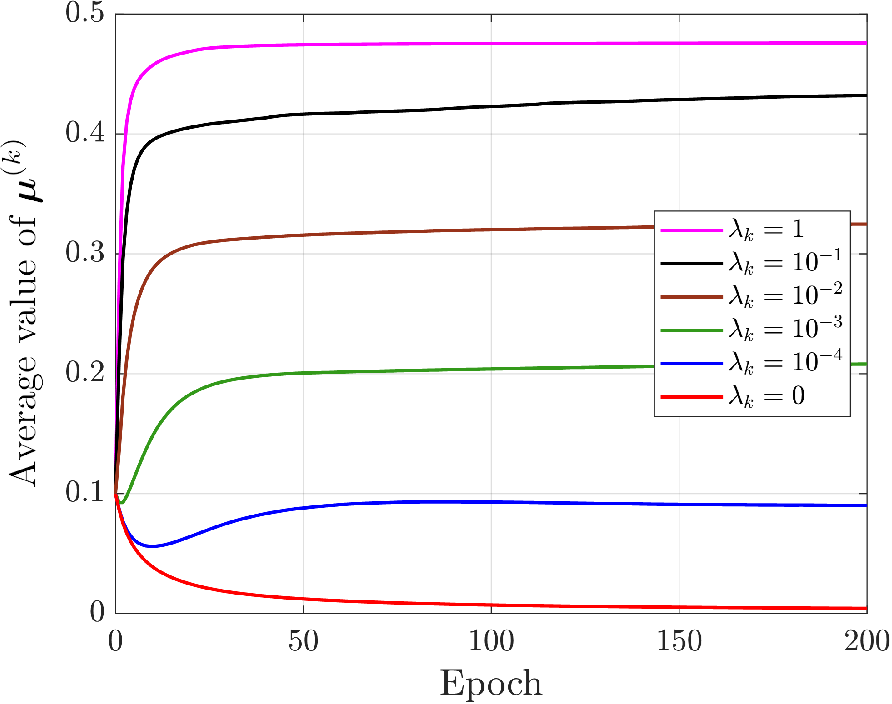,width=6cm}}\vspace{-2mm}
    \caption{Convergence behavior of the average value of bit-flip probabilities ${\bm \mu}^{(k)}$ for different regularization weight $\lambda_k$.}  \vspace{-4mm}
    \label{fig:lambda}
\end{figure}

\vspace{1mm}
{\bf Remark 2 (Impact of the Regularization Weight $\lambda_k$ on ${\bm \mu}^{(k)}$):} We analyze the convergence behavior of the average value of ${\bm \mu}^{(k)}$ in response to changes in the regularization weight $\lambda_k$, using a numerical example. In this simulation, we consider an image reconstruction task using the MNIST dataset. All values of ${\bm \mu}^{(k)}$ are initialized to $0.1$, and the loss function in \eqref{eq:mse_l2} is employed. 
Fig.~\ref{fig:lambda} shows that as $\lambda_k$ increases, the average value of ${\bm \mu}^{(k)}$ converges to higher values. This indicates that more challenging communication environments can be effectively captured by increasing $\lambda_k$. 
Fig.~\ref{fig:lambda} also shows that when $\lambda_k=0$, the average value of ${\bm \mu}^{(k)}$ converges to a very small value, requiring impractically high transmission power or communication resources to achieve such low bit-flip probabilities. 

\section{Training-Aware Communication Strategy\\of BlindSC}\label{Sec:infer}
In this section, we present a communication strategy tailored to our blind training. The core idea is to align the actual BERs with the pre-trained bit-flip probabilities to ensure reliable task performance. Specifically, we focus on a communication system that employs uncoded QAM, dynamically adjusting power and modulation levels while adhering to total power and transmission rate constraints. This system can be extended to various digital communication systems. 






\begin{figure*}[t]
    \centering
    {\epsfig{file=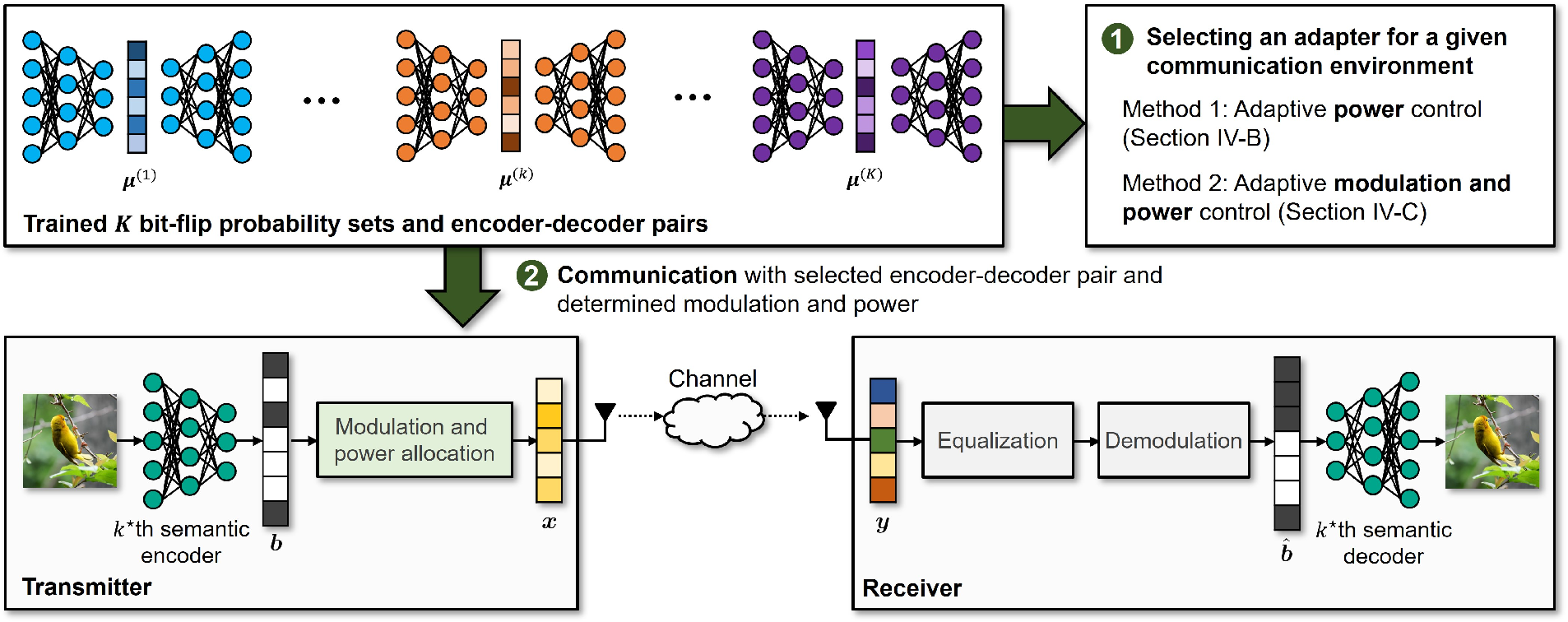,width=13cm}}\vspace{-2mm}
    \caption{The proposed training-aware communication strategy in BlindSC with two adaptive control methods.}\vspace{-3mm}
    \label{fig:Commun}
\end{figure*}


\subsection{Overview} 
Suppose that the SC system has the total transmission power constraint, $P_{\rm tot}$, and the target transmission rate, $R_{\rm target}$.  
Then, the transmitter and receiver select the most suitable encoder-decoder pair from the $K$ pre-trained pairs, while determining the transmission power and modulation level for each symbol based on the pre-trained $K$ bit-flip probability sets and the current channel-gain-to-noise-power ratio, $\gamma$. This channel-adaptive selection and optimization process is denoted as
\begin{align}\label{eq:adapt_fn}
    \{k^\star,\{p_t, m_t\}_{t=1}^{T}\} = f_{\rm adapt}(P_{\rm tot}, R_{\rm target},\{{\bm \mu}^{(k)}\}_{k=1}^K, \gamma),
\end{align}
where $k^\star$ is the index of the selected encoder-decoder pair, and $f_{\rm adapt}$ is the adaptation function, which will be detailed in Sec.~\ref{Sec:APC} and Sec.~\ref{Sec:AMPC}. Note that the function $f_{\rm adapt}$ needs to satisfy the transmission rate in \eqref{eq:Trans_rate_constraint} and the total power constraint in \eqref{eq:Total_power}. Moreover, this function should satisfy the BER matching condition that the BER, achieved through the transmission power $p_t$ and modulation level $m_t$ of the $t$th symbol carrying the $n$th bit, closely matches the $n$th element of ${\bm \mu}^{(k^\star)}$, as follows:
\begin{align}\label{eq:condition}
     \mu_n^{(k^\star)} &\approx \frac{\sqrt{2^{m_t}} - 1}{\sqrt{2^{m_t}} \log_2 \sqrt{2^{m_t}}} {\rm erfc} \left( \sqrt{\frac{3 {p_t\gamma}}{2(2^{m_t} - 1)}} \right) \nonumber\\ &~~~+ \frac{\sqrt{2^{m_t}} - 2}{\sqrt{2^{m_t}} \log_2 \sqrt{2^{m_t}}} {\rm erfc} \left( 3\sqrt{\frac{3 { p_t\gamma}}{2(2^{m_t} - 1)}} \right) \nonumber \\
     &\triangleq {\rm BER}(p_t,m_t,\gamma), 
\end{align}
where the approximation follows from \eqref{eq:mu_example}.
This condition helps minimize the discrepancy between the bit error rates observed during training and those encountered during communication, leading to improved reliability and task performance. The optimization problem solved by $f_{\rm adapt}$ is summarized as follows:
\begin{align}
    ({\bf P0})~~(k^\star, \{p_t, &m_t\}_{t=1}^T)=\!\!\!\!\argmin_{ k, \{{p}_t^{(k)}, m_t^{(k)}\}_{t=1}^{T}}\!\!\!\!\mathbb{E}[d({\bm u},\hat{\bm u})], \\[5pt]
    ~\text{s.t. } 
    &\big|{\rm BER}({p}_t^{(k)},{m}_t^{(k)},\gamma) - \mu_n^{(k)}\big|\leq \epsilon, \nonumber \\
    &\sum_{t=1}^T p_t^{(k)} \leq P_{\rm tot},~R \geq R_{\rm target},  \nonumber
\end{align}
where $p_t^{(k)}$ and $m_t^{(k)}$ represent the power and modulation level for the $t$th symbol in the $k$th bit-flip probability set, and $\epsilon \ll 1$. Note that $d({\bm u}, \hat{\bm u})$ represents the distortion measure between the input image ${\bm u}$ and its reconstruction $\hat{\bm u}$, used during training, and it decreases as $k$ increases due to the trade-off discussed in Sec.~\ref{Sec:loss}. 

Once the process above is completed, the transmitter encodes the image data using the selected $k^\star$th encoder, and performs the quantization process to obtain the bit sequence ${\bm b}$.
Then, the transmitter transmits ${\bm b}$ using the specified modulation levels and transmission powers. It should be noted that the process in \eqref{eq:adapt_fn} is also executed at the receiver. Consequently, with the knowledge of $\{p_t, m_t\}_{t=1}^{T}$, the receiver can perform the detection to obtain the estimated bit sequence $\hat{\bm b}$. 

Meanwhile, in this work, we consider an digital communication system that employs uncoded QAM to simplify the transmission and reception processes. However, the framework can be extended to incorporate advanced coding schemes by replacing the current BER function in \eqref{eq:condition} with a BER function that accounts for channel coding.
In the remainder of this section, we first introduce the adaptive power control (APC) method (see Sec.~\ref{Sec:APC}) and then explore a more advanced adaptive modulation and power control (AMPC) method (see. Sec.~\ref{Sec:AMPC}). The overall procedure of our communication strategy is illustrated in Fig.~\ref{fig:Commun}.



\subsection{APC Method}\label{Sec:APC}
The APC method in our communication strategy involves fixing the modulation level and determining the transmission power for each symbol. 
Specifically, the modulation level is set to the lowest level that satisfies the transmission rate constraint, as follows:
\begin{align}\label{eq:m_t_init}
    m_t^{(k)} &= \min\left\{m:\left\lceil \frac{NB}{m} \right\rceil \leq \frac{NB}{R_{\rm target}},m\in\{2,4,6,\cdots\}\right\} \nonumber \\
    &\triangleq m_{\rm APC},
\end{align}
for all $t,k$, where $\left\lceil \frac{NB}{m} \right\rceil$ represents the length of the symbol sequence $T$ when using $2^m$-QAM, and $\frac{NB}{R_{\rm target}}$ indicates the minimum required length of the symbol sequence to achieve the target transmission rate ${R_{\rm target}}$. 
By setting the modulation level to the minimum value, we reduce the required transmission power by leveraging the fact that higher modulation levels require greater transmission power to satisfy the BER condition in \eqref{eq:condition}. 
Additionally, this setting simplifies the problem ${\bf P0}$ by ensuring that the constraint $R \geq R_{\rm target}$ is always satisfied. In this context, the APC method solves the following optimization problem:
\begin{align}
    ({\bf P1})~~&(k^\star,\{p_t\}_{t=1}^T)=\!\!\argmin_{ k, \{{p}_t^{(k)}\}_{t=1}^{T}}\!\!\mathbb{E}[d({\bm u},\hat{\bm u})], \\[-5pt]
    &\text{s.t. } 
    \big|{\rm BER}({p}_t^{(k)},{m}_{\rm APC},\gamma) - \mu_n^{(k)}\big|\leq \epsilon,~\sum_{t=1}^T p_t^{(k)} \leq P_{\rm tot}. \nonumber 
\end{align}
To solve the problem ${\bf P1}$, the APC method first sorts the bit-flip probabilities $\{\mu_n^{(k)}\}_{\forall n}$ in ascending order and divides them into $T=\left\lceil \frac{NB}{m_{\rm APC}} \right\rceil$ groups. Next, the method computes the average bit-flip probability for each group as follows:
\begin{align}\label{eq:BER_avg}
    \bar{\mu}_t^{(k)} = \frac{1}{{m_{\rm APC}}}\sum_{n\in\mathcal{K}_t}\mu_n^{(k)}, 
\end{align}
for $t\in\{1,\cdots,T\}$, where $\mathcal{K}_t$ is the index set of the $t$th group. The required transmission power to achieve $\bar{\mu}_t^{(k)}$, denoted as $p_t^{(k)}$, is then determined by solving the equation below:
\begin{align}\label{eq:BER_solve_fn}
    \bar{\mu}_t^{(k)} = {\rm BER}(p_t^{(k)},m_{\rm APC},\gamma).
\end{align}
In general, since the BER function is monotonically decreasing with respect to $p_t^{(k)}$ for a given $m_{\rm APC}$, one can readily obtain the solution using various numerical methods such as bisection or Newton-Raphson algorithms \cite{suli2003introduction}. Further, it should be noted that by sorting $\mu_n^{(k)}$ and averaging similar values, we can assume that $\bar{\mu}_t^{(k)}\approx \mu_n^{(k)}, \forall n \in \mathcal{K}_t$; thereby, the condition in \eqref{eq:condition} is satisfied by solving the equation of \eqref{eq:BER_solve_fn}. 
After computing the power $p_t^{(k)}$ for each $t$ and $k$, the required total power for the $k$th encoder-decoder pair is obtained as
\begin{align}
    P_{\rm sum}^{(k)} = \sum_{t=1}^T p_t^{(k)}.
\end{align}
Then, to achieve the highest performance while satisfying the power constraint, the index of the most suitable encoder-decoder pair is determined as
\begin{align}\label{eq:k_star}
    k^{\star} = \min\{k:P_{\rm sum}^{(k)}\leq P_{\rm tot}\}.
\end{align}
The corresponding transmission power is set to $p_t=p_t^{(k^\star)}$. 

After determining $k^\star$, the transmitted bits are sorted to match the order of their corresponding $\mu_n^{(k^\star)}$ values, and then modulation and power allocation are performed using $m_t$ and $p_t$. On the receiver-side, the estimated bits are reordered to their original positions following the demodulation process. Note that these ordering procedures do not incur any additional communication overhead because the sorting operation for $\mu_n^{(k^\star)}$ can be performed locally at the transmitter and receiver.


A key feature of the APC method is its simplicity and efficiency in determining modulation levels and transmission powers without complex optimization processes. 
However, by fixing the modulation level uniformly across all symbols, the method fails to take the performance gains that can be obtained through jointly optimizing the modulation level and transmission power. Consequently, the overall communication efficiency and performance remain constrained. 
This limitation motivates us to develop a more advanced adaptation method, namely the AMPC method, described in the following subsection.  






\subsection{AMPC Method}\label{Sec:AMPC}

A primary goal of the proposed AMPC method is to minimize the total transmission power over the APC method, while strictly adhering to the transmission rate constraint and BER matching condition. To achieve this goal, our method aims to find the optimal modulation level and corresponding transmission power for each symbol. The corresponding optimization problem is expressed as 
\begin{align}
    ({\bf P2})~~&(k^\star,\{p_t,m_t\}_{t=1}^T)\nonumber\\
    &=\argmin_{ k, \{{p}_t^{(k)}, m_t^{(k)}\}_{t=1}^{T}}\!\!\!\!\mathbb{E}[d({\bm u},\hat{\bm u})] + \nu \sum_{t=1}^T p_t^{(k)}, \\[5pt]
    &~~~~~~~\text{s.t. } 
    \big|{\rm BER}({p}_t^{(k)},{m}_t^{(k)},\gamma) - \mu_n^{(k)}\big|\leq \epsilon, \nonumber \\
    &~~~~~~~~~~~\sum_{t=1}^T p_t^{(k)} \leq P_{\rm tot},~R \geq R_{\rm target},  \nonumber
\end{align}
where $\nu>0$ is a small constant that ensures $\mathbb{E}[d({\bm u},\hat{\bm u})] \gg \nu \sum_{t=1}^T p_t^{(k)}, \forall k$, emphasizing the importance of minimizing distortion while still considering power consumption. To explore this problem more clearly, let us consider the example where the bit sequence length is $16$ and the desired symbol sequence length is $4$, implying the $R_{\rm target}=4$. In this example, the potential combinations of $\{m_1^{(k)},m_2^{(k)},m_3^{(k)},m_4^{(k)}\}$ are $\{8,4,2,2\}$, $\{6,6,2,2\}$, $\{6,4,4,2\}$, $\{4,4,4,4\}$, along with all permutations of these elements. 
Then, in this example, the optimal combination can be determined by evaluating both the total transmission power and the corresponding distortion.

The main challenge in solving the optimization problem ${\bf P2}$ lies in its non-convexity, which arises from the discrete nature of the modulation level $m_n$ and the non-convexity of the BER function with respect to two variables $p_n$ and $m_n$. Furthermore, due to the vast number of possible combinations of modulation levels and powers, the domain space can be significantly large. As a result, traditional optimization techniques, and even exhaustive search methods, are highly impractical for finding the optimal solution. To address these challenges, we propose the AMPC method that iteratively optimizes the transmission power and modulation level for each symbol, effectively navigating the complex search space to find near-optimal solutions.

\begin{algorithm}[t]
    \setstretch{1.0}
    \caption{Adaptive Modulation and Power Control (AMPC) Method}\label{alg:AMPC}
    {\small
    {\begin{algorithmic}[1]
            \STATE Set the initial value of $m_t$, denoted as $m_{\rm init}$, from \eqref{eq:m_t_init}, $\forall t$.
            \STATE Sort $\mu_n$ in ascending order.
            \STATE ${\ell}_{\rm s} = 1$, ${\ell}_{\rm e} = {m_{\rm init}}$.
            \STATE ${\rm h}_{\rm s} = NB-{m_{\rm init}}-1$, and ${\rm h}_{\rm e} = NB$.
            \STATE $t=1$
            \WHILE{$t\leq \lfloor\frac{T}{2}\rfloor$}
            \STATE $m_{\ell} = {\ell}_{\rm e} -{\ell}_{\rm s} +1$ and $m_{{\rm h}} = {\rm h}_{\rm e} - {\rm h}_{\rm s} +1$.
            \STATE $m_{\ell}^\prime = m_{\ell}-2$ and $m_{{\rm h}}^\prime = m_{{\rm h}}+2$.
            \STATE $\bar{\mu}_{\ell} = \frac{1}{{m_{\ell}}}\sum_{n={\ell}_{\rm s}}^{n={\ell}_{\rm e}}\mu_n$ and $\bar{\mu}_{{\rm h}  } = \frac{1}{{m_{\rm h}}}\sum_{n={\rm h}_{\rm s}}^{n={\rm h}_{\rm e}} \mu_n$.
            \STATE $\bar{\mu}_{\ell}^\prime = \frac{1}{m_{\ell}^\prime}\sum_{n={\ell}_{\rm s}}^{n={\ell}_{\rm e}-2} \mu_n$ and $\bar{\mu}_{{\rm h} }^\prime = \frac{1}{m_{{\rm h}}^\prime}\sum_{n={\rm h}_{\rm s}-2}^{n={\rm h}_{\rm e}} \mu_n$.
            \STATE Compute powers ${p}_{\ell}$, ${p}_{\ell}^\prime$, ${p}_{{\rm h}}$, ${p}_{{\rm h}}^\prime$ by solving $\bar{\mu}_{\ell} = {\rm BER}({p}_{\ell},m_{\ell},\gamma)$, $\bar{\mu}_{\ell}^\prime = {\rm BER}({p}_{\ell}^\prime,m_{\ell},\gamma)$, $\bar{\mu}_{\rm h} = {\rm BER}({p}_{\rm h},m_{\rm h},\gamma)$, $\bar{\mu}_{\rm h}^\prime = {\rm BER}({p}_{\rm h}^\prime,m_{\rm h}^\prime,\gamma)$. 
            \STATE ${p}_{\rm sum} = {p}_{\ell } + {p}_{{\rm h}}$ and ${p}_{\rm sum}^\prime = {p}_{\ell }^\prime + {p}_{{\rm h}}^\prime$. 
            \IF{(${p}_{\rm sum}^\prime > {p}_{\rm sum}$) or ($m_{\ell} = m_{\rm min}$) or ($m_{\rm h} = m_{\rm max}$)} 
                \STATE $m_t = m_{\ell}$ and $m_{T-t+1} = m_{\rm h}$. 
                \STATE $p_t = {p}_{\ell}$ and $p_{T-t+1} = {p}_{\rm h}$. 
                \STATE ${\ell}_{\rm s} = {\ell}_{\rm e}+1$ and ${\rm h}_{\rm e} = {\rm h}_{\rm s}-1$.
                \STATE ${\ell}_{\rm e} = {\ell}_{\rm s} + m_{\rm init} - 1$ and ${\rm h}_{\rm s} = {\rm h}_{\rm e}-m_{\rm init}+1$. 
                \STATE $t=t+1$.
                \STATE {\bf continue}.
            \ENDIF
            \IF{${p}_{\rm sum}^\prime \leq {p}_{\rm sum}$}
                \STATE $m_t = m_{\ell}-2$ and $m_{T-t+1} = m_{\rm h}+2$. 
                \STATE $p_t = {p}_{\ell}^\prime$ and $p_{T-t+1} = {p}_{\rm h}^\prime$. 
                \STATE ${\ell}_{\rm e} = {\ell}_{\rm e}-2$ and ${\rm h}_{\rm s} = {\rm h}_{\rm s}-2$. 
            \ENDIF
            \ENDWHILE
    \end{algorithmic}}}
\end{algorithm}




\setlength{\textfloatsep}{7pt}
\begin{figure}[t]
    \centering
    {\epsfig{file=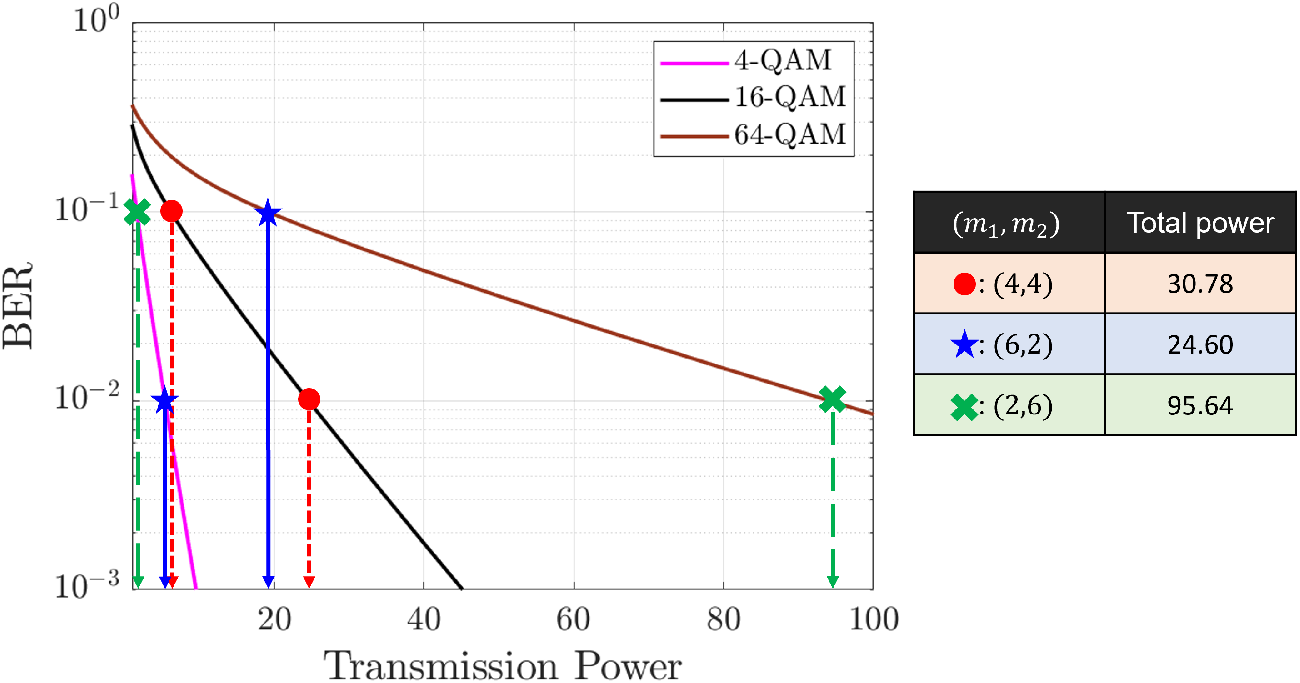,width=9cm}}\vspace{-2mm}
    \caption{Comparison of total power for various combinations of modulation levels.}
    \label{fig:motivation}
\end{figure}


The fundamental idea of the AMPC method is to assign a low modulation level to bits with low bit-flip probabilities, and a high modulation level to bits with higher bit-flip probabilities. 
The effectiveness of this idea can be verified through a simple example in which two bits need to be transmitted with bit-flip probabilities $\mu_1=10^{-1}$ and $\mu_2=10^{-2}$, respectively, and the channel-gain-to-noise-power ratio $\gamma$ is set to be one. 
Then, we examine three combinations of modulation levels $\{m_1,m_2\}$: (i) $\{4, 4\}$, (ii) $\{6,2\}$, and (iii) $\{2,6\}$, each achieving the same transmission rate. Fig.~\ref{fig:motivation} demonstrates the effectiveness of our idea, where the second combination achieves the lowest total transmission power. 
Furthermore, based on this result, it can be naturally inferred that as the difference between $\mu_1$ and $\mu_2$ increases, the total power gap between the first combination and the others also widens. 
The proposed AMPC method, motivated by the key observation mentioned above, is summarized in {\bf Algorithm~\ref{alg:AMPC}}, where the index $k$ is omitted for the sake of simplicity. The major steps of the algorithm are elaborated below. 
In Step 1, the initial modulation level is set to the same as in the APC method, ensuring the transmission rate constraint. In Step~3, the starting and ending indices for the group with the smallest bit-flip probabilities are determined. Similarly, in Step~4, the indices for the group with the highest bit-flip probabilities are determined. In the subsequent steps, the modulation levels of both groups are adjusted to minimize the total transmission power, as discussed in the earlier example. Specifically, in Step~7, the modulation levels corresponding to the low and high bit-flip probabilities, denoted as $m_{\ell}$ and $m_{\rm h}$, are calculated. In Step~8, the adjusted modulation levels are computed by decreasing $m_{\ell}$ and increasing $m_{\rm h}$. In Steps 9--12, the transmission power for each modulation level is determined by solving the BER equation, as done in the APC method. In Steps 13--25, the algorithm compares the total power of both groups to decide whether to adjust the modulation levels or keep them unchanged. If the decision is to maintain the current levels, the algorithm proceeds to the next groups. Here, the conditions for moving to the next groups are if adjusting the modulation levels increases the total power or if the modulation levels have reached their minimum or maximum values, denoted as $m_{\rm min}$ and $m_{\rm max}$.

In the AMPC method, the procedure for determining the index of the most suitable encoder-decoder pair is the same as that in equation \eqref{eq:k_star} of the APC method. Additionally, the optimized modulation level and power can be shared without any communication overhead because the AMPC method can be locally performed at the transmitter and receiver based on the pre-shared knowledge of $\{{\bm \mu}^{(k)}\}_{\forall k}$.

The main advantage of the AMPC method is its ability to consistently achieve the same or lower total transmission power compared to the APC method. This is because the AMPC method starts with the same initial modulation level as the simple method, and then adjusts the modulation level toward minimizing power consumption. This advantage will be further validated through numerical simulations in Sec.~\ref{Sec:Simul}. 

\vspace{1mm}
{\bf Remark~3 (Comparison with Conventional Modulation and Power Control Method):} In traditional digital communication systems, modulation and power control methods are widely adopted to maximize spectral efficiency or minimize BER under varying channel conditions. 
One well-known method is the water-filling algorithm, which allocates more power to channels with higher SNR and less to those with lower SNR, maximizing spectral efficiency. Unlike these conventional methods, our framework differentiates itself by dynamically assigning different modulation and power to individual data bits, even when the channel conditions remain the same for all bits. This difference arises from the fact that conventional methods typically treat all data bits equally, without accounting for the varying levels of importance or sensitivity of each bit, whereas our framework recognizes that certain bits may be more critical for performing the receiver's task. The performance gain of our strategy will be numerically demonstrated in Sec.~\ref{Sec:Simul}.

\begin{figure*}[t]
    \centering 
    \subfigure[MNIST]
    {\epsfig{file=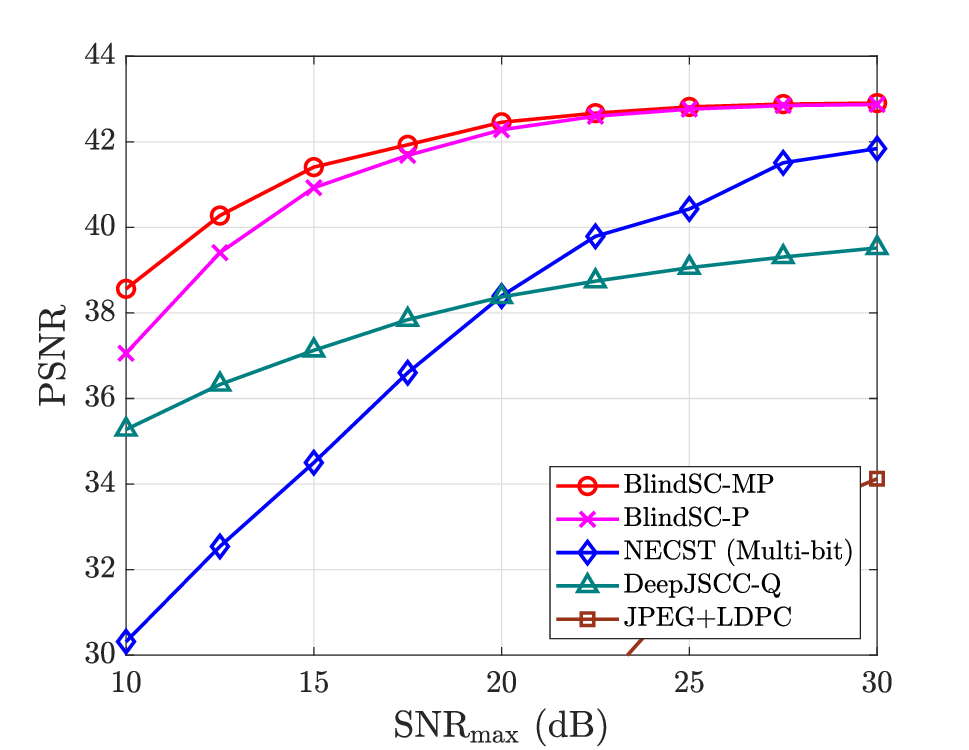, width=6cm}}
    \subfigure[CIFAR-10]
    {\epsfig{file=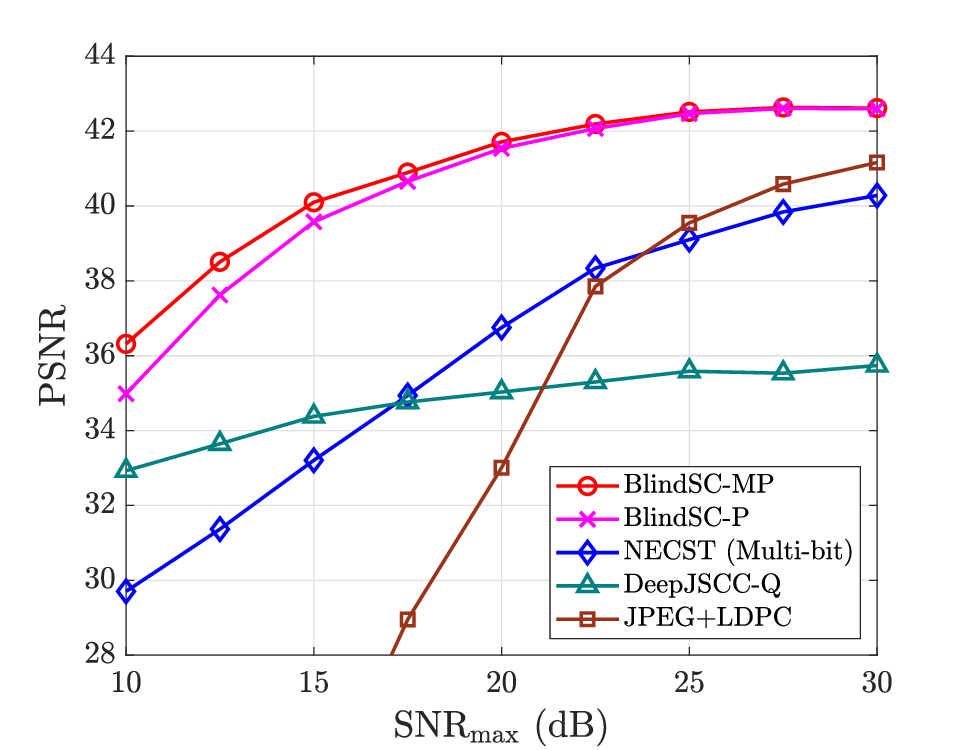, width=6cm}}
    \subfigure[STL-10]
    {\epsfig{file=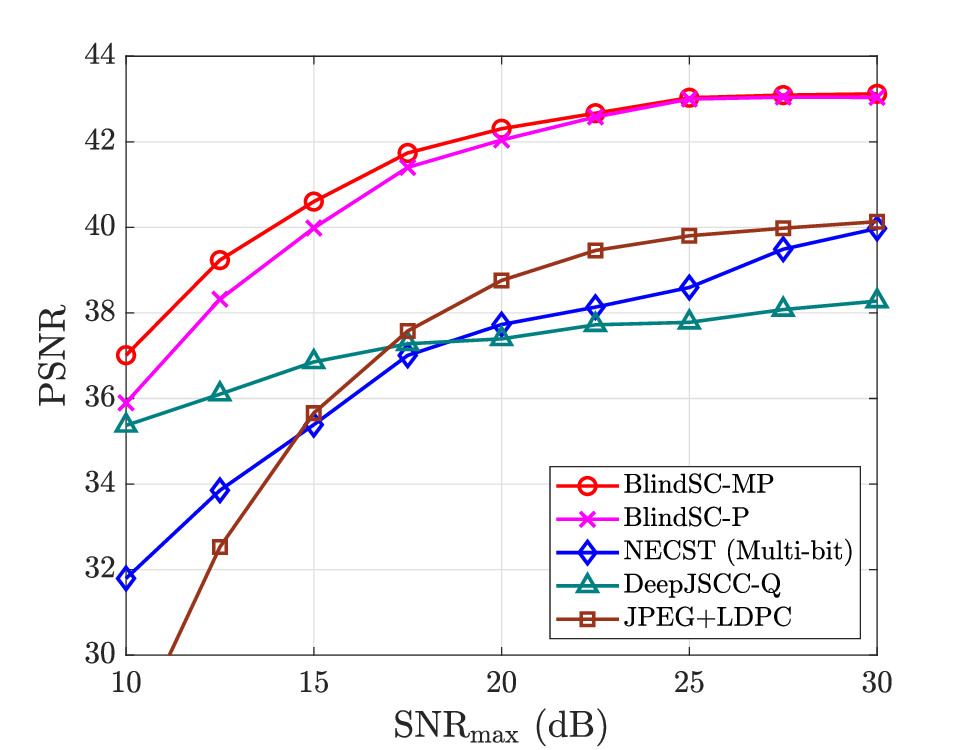, width=6cm}}\vspace{-1mm}
        \caption{Comparison of the PSNRs for various semantic and traditional communication frameworks using MNIST, CIFAR-10, and STL-10 datasets.}\vspace{-3mm}
    \label{fig:PSNR}
\end{figure*}

\section{Simulation Results and Analysis}\label{Sec:Simul}
In this section, we demonstrate the superiority of BlindSC, using the MNIST \cite{MNIST}, CIFAR-10 \cite{CIFAR10}, and STL-10 \cite{STL10} datasets. 
The MNIST dataset consists of $70,000$ grayscale images of handwritten digits, each of size $28 \times 28$ pixels, split into $60,000$ training samples and $10,000$ test samples. The CIFAR-10 dataset consists of $60,000$ color images of size $32 \times 32 \times 3$ pixels, split into $50,000$ training samples and $10,000$ test samples. The STL-10 dataset consists of $113,000$ color images of size $96 \times 96 \times 3$ pixels, split into $105,000$ training samples (including both labeled and unlabeled data) and $8,000$ test samples. 

Table~\ref{table:model} summarizes the semantic encoder and decoder architectures for each dataset, where the model architecture has been modified from that of \cite{DeepJSCC} to improve task performance. In this table, C($c$,$k$,$s$,$p$) represents a 2D convolutional layer with $c$ output channels, a $k \times k$ kernel, a stride of $s$, and padding of $p$. Similarly, CT($c$,$k$,$s$,$p$,$p_{\rm o}$) denotes a transposed 2D convolutional layer with $c$ output channels, a $k \times k$ kernel, a stride of $s$, padding of $p$, and output padding of $p_{\rm o}$. 
In the training process, the number of epochs is set to $50$, $100$, and $200$ for the MNIST, CIFAR-10, and STL-10 datasets, respectively. For all datasets, the batch size is set to $16$, and the Adam optimizer in \cite{ADAM} is employed with an initial learning rate of $10^{-4}$ to train the semantic encoder and decoder. Unless otherwise specified, the target transmission rate $R_{\rm target}$ and the total available transmission power $P_{\rm tot}$ are set to $4$ and $100$, respectively. The communication channel for all simulations is modeled as a Rayleigh fading channel. 

To evaluate the reconstruction quality, we mainly use two key metrics: PSNR and SSIM. Accordingly, the loss function is used as the MSE loss when evaluating with the PSNR, and SSIM loss when evaluating with the SSIM metric \cite{DeepJSCC_Q}. 


\begin{table}[t]
        \renewcommand{\arraystretch}{1.1}
        \caption{The semantic encoder and decoder architectures for MNIST, CIFAR-10 and STL-10 datasets.}\label{table:model}
        \setlength{\tabcolsep}{3pt}
        \footnotesize
        \centering
        {\begin{tabular}{|c|c|c|} \hline
            \multicolumn{2}{|c|}{} & \multicolumn{1}{c|}{Layers}   \\ \hline \hline
            \multirow{4}{*}{MNIST} & \multirow{2}{*}{Encoder} & \multicolumn{1}{c|}{C(32,3,1,2), PReLU, C(64,3,1,2), PReLU,}\\  
            & & \multicolumn{1}{c|}{C(64,5,2,1), PReLU, C(8,5,2,1)} \\ \cline{2-3} 
            & \multirow{2}{*}{Decoder} & \multicolumn{1}{c|}{CT(64,5,2,1,0), PReLU, CT(64,5,2,1,0), PReLU,} \\
            & & \multicolumn{1}{c|}{CT(32,3,1,2,0), PReLU, CT(1,4,1,2,0)} \\ \hline \hline
            & \multirow{3}{*}{Encoder} & \multicolumn{1}{c|}{C(64,5,2,2), PReLU, C(128,5,2,2), PReLU,} \\
            & & \multicolumn{1}{c|}{C(128,5,1,2), PReLU, C(128,5,1,2), PReLU,} \\ 
            CIFAR-10 & & \multicolumn{1}{c|}{C(24,5,1,2)} \\\cline{2-3} 
            / STL-10 & \multirow{3}{*}{Decoder} & \multicolumn{1}{c|}{CT(128,5,1,2,0), PReLU, CT(128,5,1,2,0), PReLU,} \\
            & & \multicolumn{1}{c|}{CT(128,5,1,2,0), PReLU, CT(64,5,2,2,1), PReLU,} \\ 
            & & \multicolumn{1}{c|}{CT(3,5,2,2,1)} \\ \hline
        \end{tabular}}
    \end{table}

In our simulations, we consider the following baselines for performance comparison:
\begin{itemize}
    \item {\bf BlindSC-P / BlindSC-MP}: These are the proposed digital SC frameworks, where BlindSC-P only incorporates power control while BlindSC-MP employs both power control and adaptive modulation. Unless otherwise specified, in both methods, the number of encoder-decoder pairs is set to $K=3$, and the L2-based regularization in \eqref{eq:mse_l2} is employed with regularization weights $(\lambda_1,\lambda_2,\lambda_3) = (10^{-6},10^{-3},10^{-2})$. The bit-flip probabilities are uniformly initialized within the range of $0.01$ to $0.49$ and optimized using the Adam optimizer in \cite{ADAM} with an initial learning rate of $10^{-3}$. In this framework, if the APC and AMPC methods fail to achieve the last $K$th bit-flip probability set, mainly due to the low total power or poor channel conditions, the optimized powers are adjusted by multiplying them with an appropriate constant to satisfy the total power constraint. 

    \item {\bf NECST (Multi-bit)}: This framework extends the quantization process of the digital SC framework presented in \cite{NECST} from one-bit to multi-bit. During training, the framework employs multiple BSC in \eqref{eq:bsc}, with identical bit-flip probabilities applied to all channels. Consequently, the power and modulation levels are uniformly set across all symbols under the constraints on total power and transmission rate. 

    \item {\bf DeepJSCC-Q}: This is the digital SC framework in \cite{DeepJSCC_Q}, where the real-valued output of the encoder is mapped to discrete symbols. In this framework, the soft-to-hard quantizer described in \cite{DeepJSCC_Q} is employed. 
    To match the symbol sequence length with that of other baselines, the output length of the encoder is adjusted by modifying the number of output channels in the last layer to $32$ for the MNIST dataset and $96$ for the CIFAR-10 and STL-10 datasets, resulting in encoder outputs that are considerably longer than those of the BlindSC and NECST frameworks. 
    

    \item {\bf JPEG+LDPC}: This framework utilizes a separate source-channel coding, employing the joint photographic experts group (JPEG) for source coding and the 3/4-rate low-density parity-check (LDPC) code for channel coding. To ensure the transmission rate constraint, the modulation levels are adjusted between 2 and 10.
    Additionally, the JPEG quality settings are set to 80 for the MNIST dataset and 100 for the CIFAR-10 and STL-10 datasets, ensuring compliance with the transmission rate requirements. 

    
    
    
\end{itemize}
In the BlindSC-P, BlindSC-MP and NECST frameworks, the encoder output is processed by the ReLU6 activation function, followed by an $8$-bit uniform quantizer that discretizes values within the range of 0 to 6. 

Fig.~\ref{fig:PSNR} compares the PSNRs of various frameworks across a range of ${\rm SNR}_{\rm max}$ levels in \eqref{eq:SNR_max}, using the MNIST, CIFAR-10, and STL-10 datasets. 
In this simulation, the maximum achievable SNR, denoted as ${\rm SNR}_{\rm max}$, varies with the expected channel-gain-to-noise-power ratio, $\mathbb{E}[\gamma]$, while maintaining a fixed total available transmission power $P_{\rm tot}$ as 100\footnote{This scenario can be readily extended to various communication settings by multiplying $P_{\rm tot}$ and $\mathbb{E}[\gamma]$ by the constants $c>0$ and $\frac{1}{c}$, respectively.}. Fig.~\ref{fig:PSNR} shows that BlindSC-P and BlindSC-MP frameworks consistently achieves the highest PSNR across all ${\rm SNR}_{\rm max}$ levels. 
Notably, BlindSC-P and BlindSC-MP achieve these results using only $K=3$ encoder-decoder pairs, while other frameworks utilize $9$ pairs trained for distinct ${\rm SNR}_{\rm max}$ levels. 
The performance comparison clearly indicates that BlindSC-P and BlindSC-MP consistently outperform the NECST framework, highlighting the superior effectiveness of the proposed error-adaptive blind training strategy. This is because, while the NECST framework applies a rigid constraint by enforcing a uniform bit-flip probability $\mu_n$ across all bits, BlindSC-P and BlindSC-MP relax this constraint by optimizing $\mu_n$ for each bit individually. This flexibility allows BlindSC-P and BlindSC-MP to better adapt to varying bit-reliability conditions, resulting in significantly improved performance. 
Further, the performance gap between BlindSC-P and BlindSC-MP shows the effectiveness of the AMPC method. This performance gap arises from the fact that the AMPC method operates with lower power consumption compared to the APC method, allowing it to effectively select the best-performing encoder-decoder pair, thus leading to superior overall performance.
Meanwhile, the comparison with JPEG+LDPC highlights the limitations of separate source and channel coding, as it shows much lower PSNR, especially at low SNRs. 



\begin{figure}[t]
    \centering
    {\epsfig{file=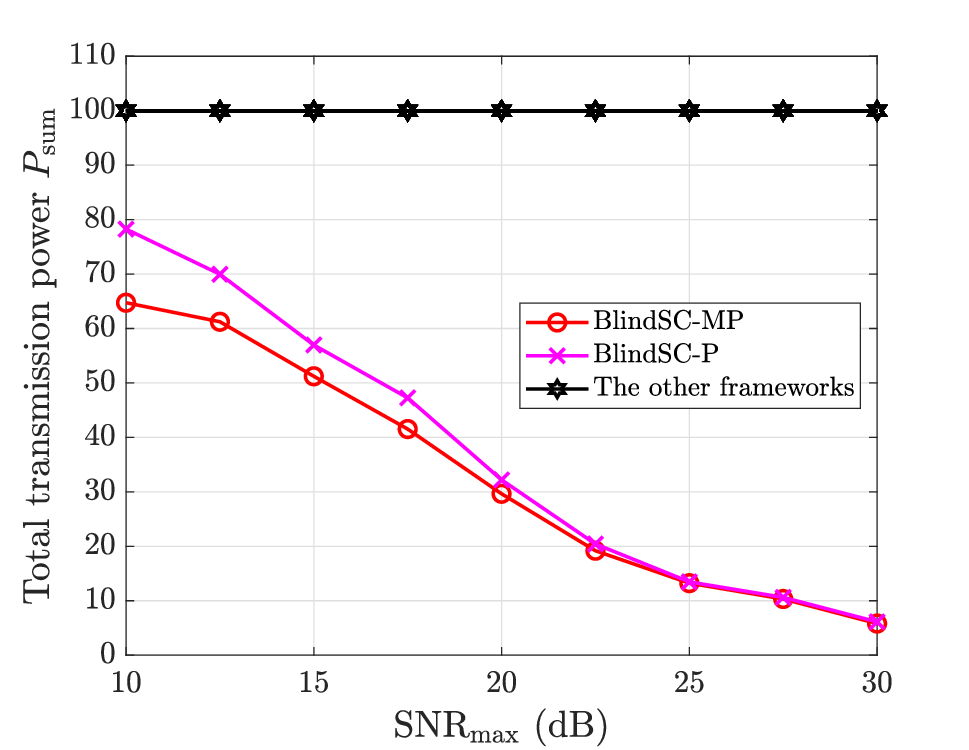,width=6.5cm}}\vspace{-2mm}
    \caption{Comparison of total transmission power between BlindSC-P, BlindSC-MP, and other frameworks at various ${\rm SNR}_{\rm max}$ levels using the CIFAR-10 dataset.}\vspace{-3mm}
    \label{fig:Total_power}
\end{figure}

Fig.~\ref{fig:Total_power} compares the total transmission power of BlindSC-P and BlindSC-MP with that of other existing frameworks across varying ${\rm SNR}_{\rm max}$ levels, using the CIFAR-10 dataset. The corresponding PSNR performance is illustrated in Fig.~\ref{fig:PSNR}(b). Note that all baseline frameworks, except BlindSC-P and BlindSC-MP, fully utilized the total available power $P_{\rm tot}$. Nevertheless, the results clearly show that BlindSC-P and BlindSC-MP not only achieve higher PSNR across all ${\rm SNR}_{\rm max}$ levels but also significantly reduce total power consumption. Within BlindSC, both the APC and AMPC methods are analyzed. The results show that the AMPC method consistently outperforms the APC method in terms of both PSNR and total transmission power, particularly at lower ${\rm SNR}_{\rm max}$ levels. This highlights the superior efficiency of the AMPC method in low-SNR communication environments, demonstrating that dynamically optimizing both transmission power and modulation levels can provide better performance than adjusting transmission power alone.




\begin{figure}[t]
    \centering
    {\epsfig{file=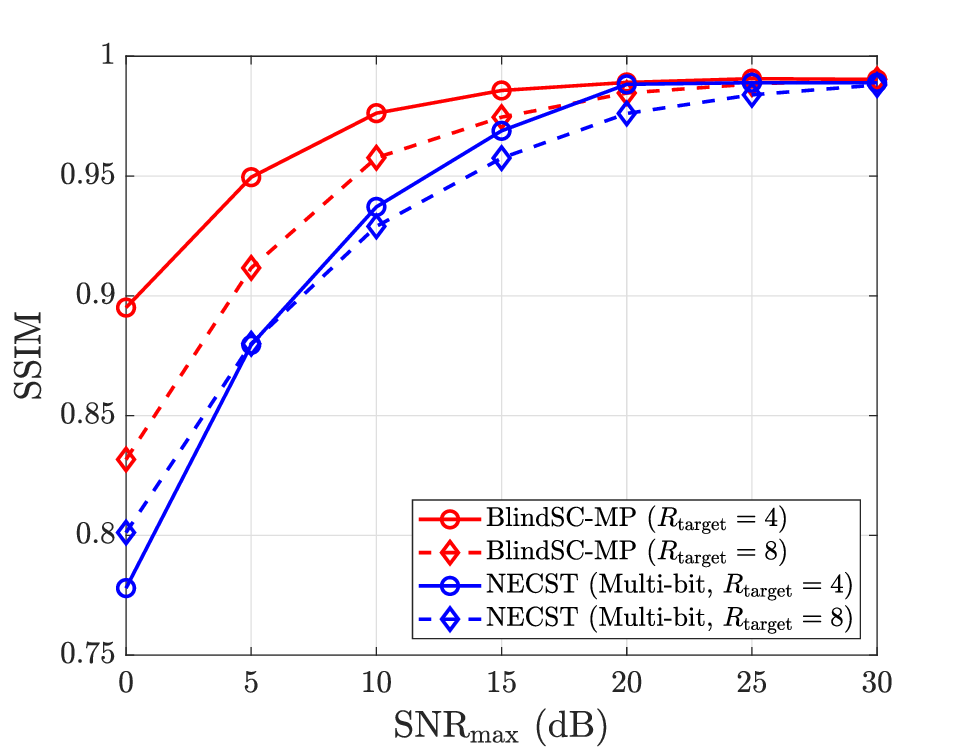,width=6.5cm}}\vspace{-2mm}
    \caption{Comparison of the SSIMs for the proposed BlindSC-MP and NECST frameworks across varying target transmission rates on the MNIST dataset.}
    \label{fig:SSIM}
\end{figure}

Fig.~\ref{fig:SSIM} compares the SSIM values of the proposed BlindSC-MP and NECST on the MNIST dataset across various target transmission rates $R_{\rm target}\in\{4,8\}$. In this comparison, NECST employs $16$-QAM and $256$-QAM to achieve $R_{\rm target}$ values of $4$ and $8$, respectively. In this simulation, as optimizing for SSIM is less challenging than optimizing for PSNR, the model was simplified by reducing the channels in the convolutional and transposed convolutional layers to 4, 8, and 4 for the original 32, 64, and 8 channels, respectively. Additionally, since the loss function was changed to optimize SSIM, the regularization weights were adjusted to $(\lambda_1,\lambda_2,\lambda_3) = (10^{-3},10^{-1},1)$. 
Fig.~\ref{fig:SSIM} demonstrates that the BlindSC-MP framework achieves the highest SSIM across varying SNRs and transmission rates. Notably, BlindSC-MP accomplishes this result with only 3 models, while NECST requires 14 models, each trained for specific combinations of ${\rm SNR}_{\rm max}$ and $R$.

\begin{figure}[t]
    \centering
    {\epsfig{file=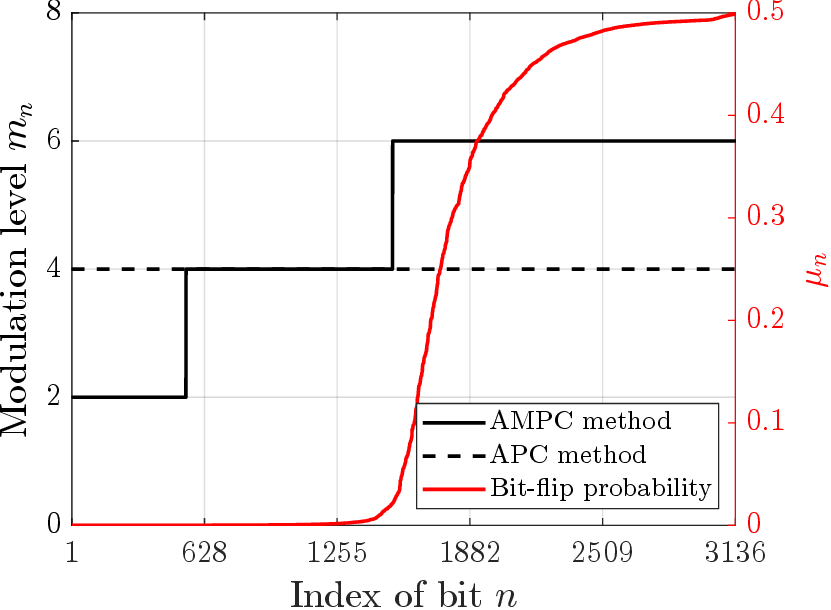,width=6.5cm}}
    \caption{Trained bit-flip probabilities and the modulation levels determined by the APC and AMPC methods on the MNIST dataset.}\vspace{-2mm}
    \label{fig:M_mu}
\end{figure}

Fig.~\ref{fig:M_mu} presents the modulation levels determined by both the APC and AMPC methods alongside the trained bit-flip probabilities. The simulation was conducted using the MNIST dataset with a trained $K$th bit probability set. Fig.~\ref{fig:M_mu} shows that the APC method applies a constant modulation level regardless of bit-flip probability. 
In contrast, in the AMPC method, as the bit-flip probability increases, the modulation level correspondingly rises. This result aligns with the core principle of the AMPC method, as described in \ref{Sec:AMPC}, where lower modulation levels are assigned to bits with lower bit-flip probabilities, and higher modulation levels are allocated to bits with higher probabilities. Consequently, by harnessing the potential performance gains from optimizing the modulation levels, the AMPC method reduces the total transmission power compared to the APC method, as shown in Figs.~\ref{fig:PSNR} and \ref{fig:Total_power}. 

\begin{figure}[t]
    \centering
    {\epsfig{file=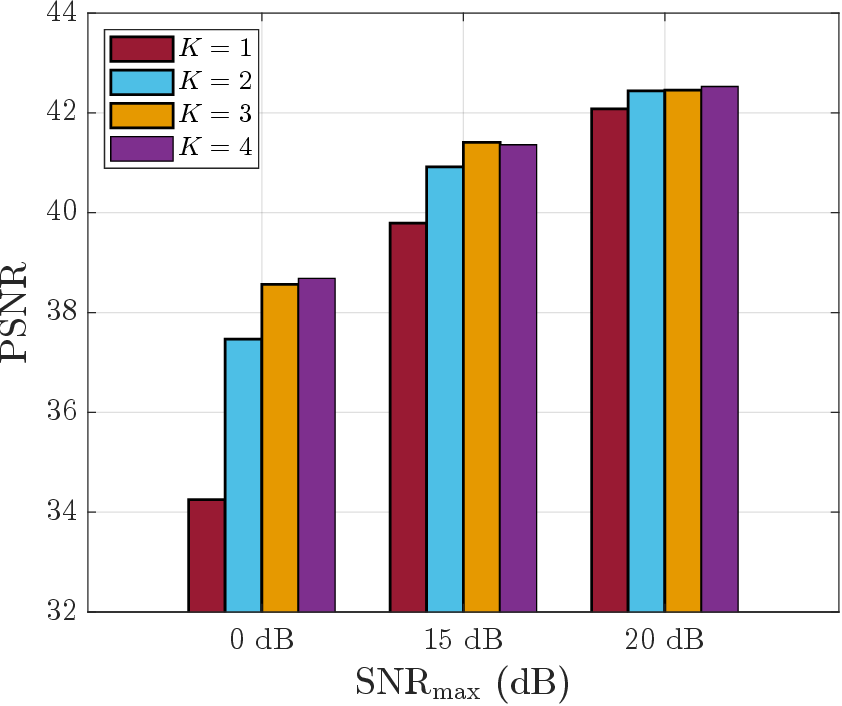,width=6cm}}
    \caption{Comparison of PSNR in BlindSC-MP for different numbers of encoder-decoder pairs $K$ across varying ${\rm SNR}_{\rm max}$ levels on the MNIST dataset.}
    \label{fig:K}
\end{figure}

In Fig.~\ref{fig:K}, the PSNR of BlindSC-MP is compared for different numbers of encoder-decoder pairs, denoted as $K$, across varying ${\rm SNR}_{\rm max}$ levels, using the MNIST dataset. In this simulation, the regularization weights are set as follows: $\lambda_1 = 10^{-6}$ for $K = 1$, $(\lambda_1, \lambda_2) = (10^{-6}, 10^{-3})$ for $K = 2$, $(\lambda_1, \lambda_2, \lambda_3) = (10^{-6}, 10^{-3}, 10^{-2})$ for $K = 3$, and $(\lambda_1, \lambda_2, \lambda_3, \lambda_4) = (10^{-6}, 10^{-4}, 10^{-3}, 10^{-2})$ for $K = 4$. Fig.~\ref{fig:K} shows that as $K$ increases, the PSNR also improves at all ${\rm SNR}_{\rm max}$ levels. This improvement is attributed to the fact that increasing $K$ enables the proposed framework to more densely represent diverse communication environments, allowing for finer adaptation to varying channel conditions. Fig.~\ref{fig:K} also illustrates that the performance difference between $K=3$ and $K=4$ is minimal. This suggests that BlindSC-MP can achieve high task performance without requiring a large number of encoder-decoder pairs, enabling more efficient deployment in practical SC systems.





\section{Conclusion}\label{Sec:Conclusion}
In this paper, we have proposed a novel digital SC framework that effectively addresses the challenges of adapting to diverse communication environments. Unlike traditional SC methods, our training strategy has eliminated the need for prior knowledge of the communication environment by treating bit-flip probabilities as trainable parameters. Additionally, our framework has introduced a training-aware communication strategy that dynamically selects the optimal encoder-decoder pair based on current channel conditions, ensuring efficient power usage while maintaining high performance. Simulation results have demonstrated that our framework significantly outperforms existing methods, achieving superior task performance, lower power consumption, and greater adaptability with fewer encoder-decoder models. 
An important direction for future research is to develop an analytical method for determining the optimal number of models and $\lambda_k$ values.  



\bibliographystyle{IEEEtran}
\bibliography{Reference}

\end{document}